\documentclass[reprint,nofootinbib,superscriptaddress,amsmath,amssymb,aps,prd]{revtex4-2}
\usepackage{amsmath,mathrsfs,amsbsy,color,graphicx,bm,amsthm,amsfonts}
\usepackage{bbm}
\usepackage{times}
\usepackage{units}
\usepackage{dcolumn}
\usepackage{graphicx}
\usepackage{epsfig}
\usepackage{epstopdf}
\usepackage[colorlinks,linkcolor=red,anchorcolor=green,citecolor=blue,CJKbookmarks=True]{hyperref}
\DeclareMathSymbol{\shortminus}{\mathbin}{AMSa}{"39}
\usepackage{mathrsfs}
\usepackage{braket}
\usepackage{amssymb}
\usepackage{txfonts}
\usepackage{float}
\usepackage{enumitem}
\usepackage{multirow}

\newcommand{\meq}[1]{(\ref{#1})}

\newcommand{\hsp}{\hspace{0.1mm}}

\newcommand{\pp}{\partial}

\begin{document}

\title{Harvesting correlations from BTZ black hole coupled to a Lorentz-violating vector field}

\author{Xiaofang Liu}
\affiliation{Department of Physics, Key Laboratory of Low Dimensional Quantum Structures and Quantum Control of Ministry of Education, and Synergetic Innovation Center for Quantum Effects and Applications, Hunan Normal
University, Changsha, Hunan 410081, P. R. China}

\author{Wentao Liu}
\email[]{wentaoliu@hunnu.edu.cn (Corresponding authors)} \affiliation{Department of Physics, Key Laboratory of Low Dimensional Quantum Structures and Quantum Control of Ministry of Education, and Synergetic Innovation Center for Quantum Effects and Applications, Hunan Normal
	University, Changsha, Hunan 410081, P. R. China}

\author{Zhilong Liu}
\affiliation{Department of Physics, Key Laboratory of Low Dimensional Quantum Structures and Quantum Control of Ministry of Education, and Synergetic Innovation Center for Quantum Effects and Applications, Hunan Normal
University, Changsha, Hunan 410081, P. R. China}

\author{Jieci Wang}
\email[]{jcwang@hunnu.edu.cn (Corresponding authors)} \affiliation{Department of Physics, Key Laboratory of Low Dimensional Quantum Structures and Quantum Control of Ministry of Education, and Synergetic Innovation Center for Quantum Effects and Applications, Hunan Normal
University, Changsha, Hunan 410081, P. R. China}

\begin{abstract}
In this paper, we investigate the effects of Lorentz violation on correlations harvesting, specifically focusing on the harvested entanglement and harvested mutual information between two Unruh-DeWitt detectors interacting with a quantum field in the Lorentz-violating BTZ-like black hole spacetime. Our findings reveal that Lorentz symmetry breaking has contrasting impacts on entanglement harvesting and mutual information harvesting in BTZ backgrounds: it enhances mutual information harvesting while suppressing entanglement harvesting. This phenomenon suggests that the increase in total correlations in Lorentz-violating vector field backgrounds with gravitational coupling is predominantly driven by classical components, with quantum correlations contributing less to the overall mutual information. These results indicate that Lorentz violation, as a quantum property of spacetime, may impose intrinsic constraints on the quantum information capacity encoded in spacetime due to competition among quantum degrees of freedom for resources. Furthermore, Lorentz symmetry breaking expands the \textit{entanglement shadow} region, further demonstrating its disruptive effect on quantum correlations.


\end{abstract}

\maketitle
\section{Introduction}


It has been pointed out that the vacuum state of a free quantum field  maximally violates Bell's inequality and contains correlations between regions separated by both time and space \cite{Summers:1985pzz,Summers:1987squ,Ng:2018ilp}. These correlations can be extracted using a pair of initially uncorrelated two-level Unruh-Dewitt (UDW) detectors that interact with the vacuum field for a period of time, known as the correlation harvesting protocol \cite{Valentini:1991eah,Reznik:2002fz,Reznik:2003mnx,Salton:2014jaa,Pozas-Kerstjens:2015gta}. 
Numerous studies have demonstrated that the efficiency of correlation harvesting in quantum entanglement is critically dependent on the detector's motion, its energy gap, and the underlying spacetime structure, which includes curvature, dimensionality, and topology \cite{VerSteeg:2007xs,Henderson:2018lcy,Ng:2018drz,Martin-Martinez:2015qwa}. 
The quantum resource harvesting protocol, originally formulated using the UDW particle detector model, has now been successfully generalized to curved spacetime scenarios \cite{Henderson:2017yuv},  solidifying its importance as a pivotal subfield within relativistic quantum information research \cite{Fuentes-Schuller:2004iaz,Ahn:2008zf,Zhou:2022nur,Zhang:2020xvo,Gallock-Yoshimura:2021yok,Cong:2018vqx,Svidzinsky:2024tjo,Kukita:2017etu,Xu:2020pbj,Tjoa:2020eqh,Foo:2020dzt,Foo:2020xqn,Zhou:2021nyv,Wu:2022lmc,Wu:2022xwy,Chen:2023xbc,Wu:2023sye,Wu:2023spa,Wu:2024pwa,Liu:2024pse,Liu:2024yrf,AraujoFilho:2024ctw,AraujoFilho:2025hkm,Tang:2025eew,Li:2025bzd}.

Inspired by the recent finding that Lorentz violation can alleviate entanglement degradation \cite{Liu:2024wpa}, we note that, when considered as a quantum feature of spacetime, such violation could play a pivotal role in the UDW detector model.
Lorentz violation would dynamically couple with detector motion, energy gap gradients, and spacetime curvature to modulate entanglement harvesting rates.
Given that black holes are often used as testbeds for fundamental quantum theories, possible strong Lorentz-violating conditions near the event horizon--conditions that cannot be replicated on Earth--offers an ideal scenario for investigating the impact of Lorentz violation on entanglement harvesting in black hole spacetimes.
Moreover, extensive studies have been conducted on both entanglement harvesting and mutual information harvesting in various black hole environments, including rotating black holes, topological black holes, geons, and scenarios involving gravitational wave effects, among others \cite{Robbins:2020jca,Henderson:2022oyd,deSouzaCampos:2020bnj,Cong:2020nec,Liu:2022uhf,Maeso-Garcia:2022uzf,Liu:2023awu,Liu:2023zro,Lindel:2023rfi,Ji:2024fcq,Wu:2024whx,Zeng:2024qme,Wu:2024qhd,Naeem:2022drs,Simidzija:2018ddw,Bueley:2022ple,Ghosh:2023xes,Rubio:2024ryv}.

On the other hand, Lorentz invariance, as a fundamental symmetry, plays a crucial role in quantum field theory. However, the development of unified canonical theories and the observation of high-energy cosmic ray signals  \cite{ Takeda:1998ps}, suggest that spontaneous Lorentz symmetry breaking may occur at higher energy scales. 
In general, Lorentz violation effects can only be observed empirically at sufficiently low energy scales \cite{Casana:2017jkc}, and such effects can be characterized using effective field theory \cite{Kostelecky:1994rn}. 
The bumblebee gravity theory is a simple and effective classical field theory model for studying Lorentz violation \cite{Kostelecky1989,Kostelecky:2003fs,Bluhm:2004ep}.  
In this model, the introduction of the bumblebee vector field $B_{\mu}$ with a nonzero vacuum expectation value (VEV) leads to the spontaneous breaking of Lorentz symmetry, implying that the geometric structure of the background spacetime is no longer completely symmetric. 
Consequently, the bumblebee gravity theory can unveil previously unknown physical phenomena, making it critical for the evolution of modern physics \cite{Casana2018,Ovgun2019,Gullu2020,Poulis:2021nqh,Maluf2021,Xu:2022frb,Ding2022,Filho:2022yrk,Liu:2022dcn,Mai:2023ggs,Xu:2023xqh,Zhang:2023wwk,Lin:2023foj,Chen:2023cjd,Chen:2020qyp,Singh:2024nvx,Liu:2024oeq,Mai:2024lgk,Ge:2025xuy,EslamPanah:2025zcm}.

Given that Lorentz violation modify the background spacetime geometry, in line with our original motivation, we investigate how a Lorentz-violating vector field in the BTZ-like black hole spacetime influences the extraction of vacuum correlations by UDW detectors. To address this question, we compute entanglement harvesting and mutual information harvesting as functions of various physical parameters. Subsequently, we demonstrate the effects of Lorentz violation on harvested correlations within BTZ-like black hole using numerical calculations.
In this paper, we investigate both entanglement harvesting and mutual information harvesting. Mutual information serves as a metric that quantifies the total sum of classical and quantum correlations, including entanglement. Building on these measurements, we further explore how Lorentz violation, as a quantum property of spacetime, differentially affects classical and quantum correlations.



This paper is organized as follows: In Sec \ref{sec2}, we introduce the bumblebee gravity theory model and derive the Wightman function for Lorentz-violating BTZ-like spacetime. 
In Sec \ref{sec3}, we give the expressions for the correlations in the interaction of the UDW detectors with the Lorentz-violating vector field. 
In Sec \ref{sec4}, we demonstrate the effect of Lorentz violation on entanglement harvesting and mutual information harvesting with the help of numerical calculations. 
Sec \ref{sec5} is the conclusion and outlook of the paper. Throughout this paper, we employ the natural units $\hbar=c=1$.

\section{Lorentz-violating BTZ black holes and Quantum fields}\label{sec2}

We present a concise overview of the Einstein-Bumblebee gravity framework, a theoretical extension of General Relativity (GR).  
The action governing the bumblebee field $B_\mu$ coupled to spacetime curvature is expressed as \cite{Casana2018}  
\begin{equation}
\begin{aligned}\label{Action}
\mathcal{S}_B=&\int d^3x \sqrt{-g}\left[\frac{1}{2\kappaup}\left(R-2\Lambda\right)+\frac{\varrho}{2\kappa} B^\mu B^\nu R_{\mu\nu}\right.\\ 
&\left.-\frac{1}{4}B^{\mu\nu}B_{\mu\nu}-V\left(B^\mu B_\mu\pm b^2\right)\right],
\end{aligned}
\end{equation}
where the gravitational coupling constant $\kappa = 8\pi G_N$ (with $G_N \equiv 1$ in natural units), $\Lambda$ denotes the cosmological constant, and $\varrho$ parametrizes the non-minimal coupling between the bumblebee field $B_\mu$ and gravity. 
The antisymmetric field strength tensor $B_{\mu\nu} \equiv \partial_\mu B_\nu - \partial_\nu B_\mu$ characterizes the bumblebee dynamics. 
Crucially, the potential $V$ enforces spontaneous Lorentz symmetry breaking via a non-zero vacuum expectation value $\langle B_\mu \rangle = b_\mu$, attaining its minimum when $B_\mu B^\mu = \mp b^2$. Here, $b \in \mathbb{R}^+$ defines the symmetry-breaking scale, with the $\pm$ sign distinguishing timelike ($+$) and spacelike ($-$) configurations of $B_\mu$.

Taking the variation of $g_{\mu\nu}$ and $B_\mu$ yields the effective gravitational equation $ \mathcal{G}_{\mu\nu}=0 $ and the bumblebee field equation $ \Pi_\mu=0 $, respectively. 
Here,
\begin{align}
\mathcal{G}_{\mu\nu}=&R_{\mu\nu}-\frac{1}{2}g_{\mu\nu}\left(R-2\Lambda\right)-\kappa T^{B}_{\mu\nu} \label{EinsteinEQ}, \\ \label{VEq}
\Pi_\mu=&\nabla^\mu B_{\mu\nu}-2V'B_\nu+\frac{\varrho}{\kappa}B^\mu R_{\mu\nu}.
\end{align}
$T^{B}_{\mu\nu}$ is the bumblebee energy momentum tensor, which have the following form:
\begin{equation}\label{TBab}
T^{B}_{\mu\nu}=B_{\mu\alpha}B^\alpha\hsp_\nu-\frac{1}{4}g_{\mu\nu}B^{\alpha\beta}B_{\alpha\beta}-g_{\mu\nu}V+2B_\mu B_\nu V'
+\frac{\varrho}{2\kappa}\mathcal{B}_{\mu\nu},
\end{equation}
with
\begin{equation}
\begin{aligned}
\mathcal{B}_{\mu\nu}=&
g_{\mu\nu}B^{\alpha}B^{\beta}R_{\alpha\beta}-4B_{(\mu} B^\alpha R_{\nu) \alpha}+\nabla_\alpha\nabla_\mu\left(B^\alpha B_\nu\right)
\\&+\nabla_\alpha\nabla_\nu\left(B^\alpha B_\mu\right)
-\nabla^2\left(B_\mu B_\nu\right)-g_{\mu\nu}\nabla_\alpha\nabla_\beta\left(B^\alpha B^\beta\right).
\end{aligned}
\end{equation}
Our goal is to explore the impact of a Lorentz-violating vector field on entanglement harvesting between two particle detectors in a BTZ-like black hole spacetime. 
To achieve this, we need to solve the field equations to obtain the background spacetime geometry affected by Lorentz violation. 
As a prerequisite for solving these equations, the specific form of the potential $V$ must be determined.
To explore the effect of Lorentz violation in (2+1)-dimensional AdS $ (\text{AdS}_3) $ spacetime with a nonzero cosmological constant, we consider the potential proposed by Maluf and Neves \cite{Maluf2021}, which allows a simple linear form:
\begin{align}
V=V(\lambda,X)=\frac{\lambda}{2} X,
\end{align}
where $ \lambda $ is interpreted as a Lagrange-multiplier field \cite{Bluhm:2007bd}.
The equation ensures that, for any on-shell field $ \lambda $ in the vacuum condition $ X=0 $, the potential $ V=0 $.
Interestingly, the potential function in the above form behaves similarly to a cosmological constant. 
This particular assumption leads us to consider:
\begin{align}\label{VVV}
&V(B^\mu B_\mu-b^2)=\frac{\lambda}{2}(B^\mu B_\mu-b^2)=0,\\ \label{Vp}
&V'(B^\mu B_\mu-b^2)=\frac{\lambda}{2},
\end{align}
where $ V'(X)=dV(X)/dX $.
Then, we assume the metric corresponds to a black hole of (2+1) dimensions and adopt the following line element:
\begin{equation}
ds^2= -A(r)dt^2+F(r)dr^2+r^2d\phi^2,
\end{equation}
where $ A(r) $ and $ F(r) $ are some undetermined functions.
The form corresponding to the above metric for the bumblebee field $ B_\mu $ is given by:
\begin{equation}
B_\mu=\left(0, b\sqrt{F(r)},0  \right),
\end{equation}
so that the constant norm condition $ B^{\mu}B_{\mu}=b^2 $ is satisfied.
Then, the nonzero components of the effective gravitational field equations and the equations of motion for the bumblebee field, both associated with the metric, are given as
\begin{align}\label{EQG1}
&\mathcal{G}_{tt}=\frac{(1+\alpha)A \pp_rF}{2r F^2}- A\Lambda,\\
&\mathcal{G}_{rr}=(\Lambda-b^2\kappa \lambda)F+\frac{(1+\alpha)\pp_rA}{2rA}+\frac{\alpha \pp_rF}{2r F}+\frac{\alpha \Upsilon }{2A},\\
&\mathcal{G}_{\phi\phi}=r^2\Lambda-\frac{(1+\alpha)r^2\Upsilon}{2AF},\\ \label{EQE2}
&\Pi_r=\frac{1}{\kappa b\sqrt{F}}\left(\frac{\alpha\pp_rF}{2rF^2}+\frac{\alpha \Upsilon}{2AF}-\kappa b^2\lambda \right).
\end{align}
where $ \alpha= \varrho b^2 $ is the Lorentz-violating parameter, and
\begin{equation*}
\Upsilon=\frac{(\pp_rA)^2}{2A}+\frac{\pp_r A\pp_r F}{2F}-\pp_r^2A.
\end{equation*}
To solve the system of differential equations composed of Eqs. \meq{EQG1}-\meq{EQE2}, we construct two linear combinations, which are given by
\begin{align*}
2r F^2\left(\mathcal{G}_{tt}+\tfrac{A}{F}\mathcal{G}_{rr}-\kappa b \sqrt{F}A \,\Pi_r \right)=0,\quad
\alpha \mathcal{G}_{\phi\phi}+\tfrac{(1+\alpha)r^2}{F}\mathcal{G}_{rr}=0,
\end{align*}
where the first equation leads to
\begin{equation*}
\pp_r(AF)=0~~~\Rightarrow~~~ F=\mathcal{C}_1/A,\, \text{with}\,\, \mathcal{C}_1\,\, \text{is constant},
\end{equation*}
and the second equation, based on the above result, becomes
\begin{equation}\label{AAAAA}
\frac{(1+\alpha)}{2\mathcal{C}_1}r\pp_rA+r^2\left[\alpha\Lambda+(1+\alpha)(\Lambda-\kappa b^2\lambda) \right]=0.
\end{equation}
By integrating Eq. \meq{AAAAA}, we obtain the specific form of the metric function $ A(r) $:
\begin{equation}
A(r)=r^2 \mathcal{C}_1\left( \kappa b^2\lambda-\frac{1+2\alpha}{1+\alpha}\Lambda \right)+\mathcal{C}_2,
\end{equation}
where $ \mathcal{C}_2 $ is an integration constant. 
At this stage, Eqs. \meq{EQG1}-\meq{EQE2} remain nonzero, sharing a common factor $ (1+\alpha)\kappa b^2\lambda-2\alpha \Lambda $.
To satisfy all the constraint equations, we assume the following relation between the Lagrange multiplier field $ \lambda $ and the cosmological constant:
\begin{equation}
\lambda:=\frac{2\alpha \Lambda}{(1+\alpha)\kappa b^2}.
\end{equation}
Subsequently, we obtain the metric function that satisfies all the constraint equations:
\begin{equation}
A(r)=-\frac{\mathcal{C}_1 \Lambda}{1+\alpha}r^2+\mathcal{C}_2.
\end{equation}
Here, we set $ \mathcal{C}_1 = (1 + \alpha) $ and consider the case of a negative cosmological constant to correspond to the asymptotic behavior of the BTZ black hole. 
In this context, we define the AdS radius, also known as the cosmological length scale, as $ \ell = \sqrt{-1/\Lambda} $.
In general, the integration constant $ \mathcal{C}_2 $ represents the mass of a BTZ black hole, and we can set $ \mathcal{C}_2 = M $. 
This leads to a Lorentz-violating BTZ-like black hole, given by
\begin{equation}\label{ds2}
g_{\mu\nu}=\text{diag}\left\{M-\frac{r^2}{\ell^2},~\frac{1+\alpha}{M-r^2/\ell^2},~r^2 \right\}.
\end{equation}
This result is consistent with Ref. \cite{Ding:2023niy} in the case of $ J=0 $.
The Kretschmann scalar is 
\begin{align}
R^{\mu\nu\sigma\tau}R_{\mu\nu\sigma\tau}=\frac{12}{\ell^4(1+\alpha)^2},
\end{align}
it is clear that the spacetime \meq{ds2} is singular as $ \alpha=-1 $.
The horizons of the black hole are located at  
\begin{align}
r_h=\ell\sqrt{M},
\end{align} 
where the horizon radius of the black hole, which does not depend on the spontaneous Lorentz symmetry breaking, is consistent with the standard BTZ black hole.
The spacetime characteristics of this class of Lorentz-violating black holes are such that they are not specially spherically symmetric but generally so (i.e., $-g_{tt}g_{rr}\neq 1$) \cite{Casana2018}, which makes their spacetime structure strongly dependent on $\alpha$.  
While rescaling and coordinate transformations can mathematically relate our metric to the standard BTZ form, the physical context and geometric properties reveal a fundamental distinction.  
The dependence on $\alpha$ is physically meaningful because it originates from the Lorentz-violating dynamics of the Einstein-Bumblebee model, leading to a spacetime that is structurally and physically unique from the standard BTZ black hole. 
Furthermore, it is noteworthy that if we apply a conformal transformation to the line element with the reparametrizations $\tilde{M} = M / \sqrt{1 + \alpha}$ and $\tilde{\ell}^2 = \sqrt{1 + \alpha} \ell^2$, we obtain a metric of the form characterized by a conical deficit
\begin{equation}
d\tilde{s}^2 = -\left(\tilde{M}-\frac{r^2}{\tilde{\ell}^2}\right)dt^2 +\left(\tilde{M}-\frac{r^2}{\tilde{\ell}^2}\right)^{-1} dr^2+ \frac{r^2}{\sqrt{1+\alpha}}d\phi^2.
\end{equation}
Thus, it appears that we can also consider the origin of the parameter $\alpha$ as a conical deficit.
On the other hand, if the integration constant $ \mathcal{C}_2=0 $, the resulting spacetime corresponds to an $ \text{AdS}_3 $ background modified by Lorentz violation:
\begin{equation}\label{etaaa}
\eta^\text{Lv}_{\mu\nu}=\text{diag}\left\{-\frac{r^2}{\ell^2},~\frac{(1+\alpha)\ell^2}{r^2},~r^2 \right\},
\end{equation}
When $ \alpha=0 $, the metric \meq{etaaa} reduces to the standard (2+1)-dimensional AdS ($ \text{AdS}_3 $) spacetime.
In fact, this metric \meq{etaaa} can also be obtained by performing a simple coordinate transformation on the standard $ \text{AdS}_3 $ spacetime and redefining the AdS radius as
\begin{equation}\label{tell}
t\rightarrow\sqrt{1+\alpha}\, t,\quad \ell \rightarrow \sqrt{1+\alpha}\, \ell,
\end{equation}
thus making explicit the role of the Lorentz-violating parameter $ \alpha $.
Meanwhile, through the coordinate transformation \meq{tell}, we can identify the two-point correlation function required for the numerical integration of entanglement harvesting in the Lorentz-violating spacetime.

The Wightman function for a conformally coupled quantum scalar field $ \hat{\phi}(\mathbf{x}) $ in the Lorentz-violating $ \text{AdS}_3 $-like spacetime is given by \cite{Henderson:2018lcy,Henderson:2019uqo}
\begin{equation}
W^\text{Lv}_{\text{AdS}_3}(\mathbf{x},\mathbf{x}')=\!\frac{\sqrt{(1+\alpha)}^{-1}}{4\pi\ell\sqrt{2}}\!
\left[\frac{1}{\!\sqrt{\sigma(\mathbf{x},\mathbf{x}')}}\!-\!\frac{\zeta}{\!\sqrt{\sigma(\mathbf{x},\mathbf{x}')+2}} \right]\!,
\end{equation}
where
\begin{equation}
\begin{aligned}
\sigma(\mathbf{x},\mathbf{x}')=&\frac{r r'}{\ell^2(1+\alpha)}\cosh\Delta\phi
-\sqrt{\tfrac{r^2-\ell^2(1+\alpha)}{\ell^2(1+\alpha)}}\\
&\times\sqrt{\tfrac{r'^2-\ell^2(1+\alpha)}{\ell^2(1+\alpha)}} 
\cosh\left(\tfrac{\sqrt{1+\alpha}\Delta t}{\ell\sqrt{1+\alpha}}\right)-1,
\end{aligned}
\end{equation}
with \(\Delta\phi := \phi - \phi'\) and \(\Delta t := t - t'\).
The parameter \(\zeta \in \{-1,0,1\}\) specifies the boundary condition for the field at the spatial infinity: 
Neumann (\(\zeta=-1\)), transparent (\(\zeta=0\)), and Dirichlet (\(\zeta=1\)).
In this paper, we will consistently adopt the Dirichlet boundary condition.
We now consider a conformally coupled quantum scalar field $ \hat{\phi}(\mathbf{x}) $ in the Lorentz-violating BTZ-like spacetime. 
By choosing the Hartle–Hawking vacuum $ \ket{0} $, we can construct the corresponding Wightman function $ W^\text{Lv}_\text{BTZ} $ through an image sum of the Wightman function in the Lorentz-violating $ \text{AdS}_3 $-like spacetime \cite{Lifschytz:1993eb}.
Let $ \Gamma: (t,r,\phi)\rightarrow(t,r,\phi+2\pi) $ represent the identification of a point $ \mathbf{x} $ in $ \text{AdS}_3^\text{Lv} $ spacetime \cite{Bueley:2022ple}.
Then,  $ W^\text{Lv}_\text{BTZ} $ is known to be 
\begin{equation}
\begin{aligned}
W^\text{Lv}_\text{BTZ}(\textbf{x},\textbf{x}')&=\sum^{\infty}_{n=-\infty}W^\text{Lv}_{\text{AdS}_3}\left(\textbf{x},\Gamma^n\textbf{x}' \right)\\
&=\frac{1}{4\pi\ell\sqrt{2(1+\alpha)}}
\sum^{\infty}_{n=-\infty}\Pi\left(\textbf{x},\Gamma^n\textbf{x}'\right),
\end{aligned}
\end{equation}
where
\begin{align}
\begin{aligned}
\Pi\left(\textbf{x},\Gamma^n\textbf{x}'\right)=\frac{1}{\sqrt{\sigma_n \left(\textbf{x},\Gamma^n\textbf{x}'\right)}}
-\frac{\zeta}{\sqrt{\sigma_n \left(\textbf{x},\Gamma^n\textbf{x}'\right)+2}},
\end{aligned}\\
\begin{aligned}
\sigma_n\left(\textbf{x},\Gamma^n\textbf{x}'\right)=&\frac{rr'}{r_h^2}\cosh\left[ \tfrac{r_h \left(\Delta\phi-2\pi n \right)}{\ell\sqrt{1+\alpha}} \right]-1\\
&-\tfrac{\sqrt{(r^2-r_h^2)({r'}^2-r^2_h)}}{r^2_h}\cosh\left(\tfrac{r_h\Delta t}{\ell^2\sqrt{1+\alpha}} \right).\label{eq28}
\end{aligned}
\end{align}
Notably, in the Wightman function, the $n=0$ term is similar to the AdS-Rindler term \cite{Jennings:2010vk}, which corresponds to a uniformly accelerating detector in the Lorentz-violating $\text{AdS}_3$-like spacetime. 
The remaining $ (n\neq0) $ terms can be referred to as the Lorentz-violating BTZ-like terms, which contribute to the black hole structure.

\begin{figure*}[t]
\centering ~\\~ ~\\~ ~\\~ ~\\~ 
\includegraphics[width=0.9\linewidth]{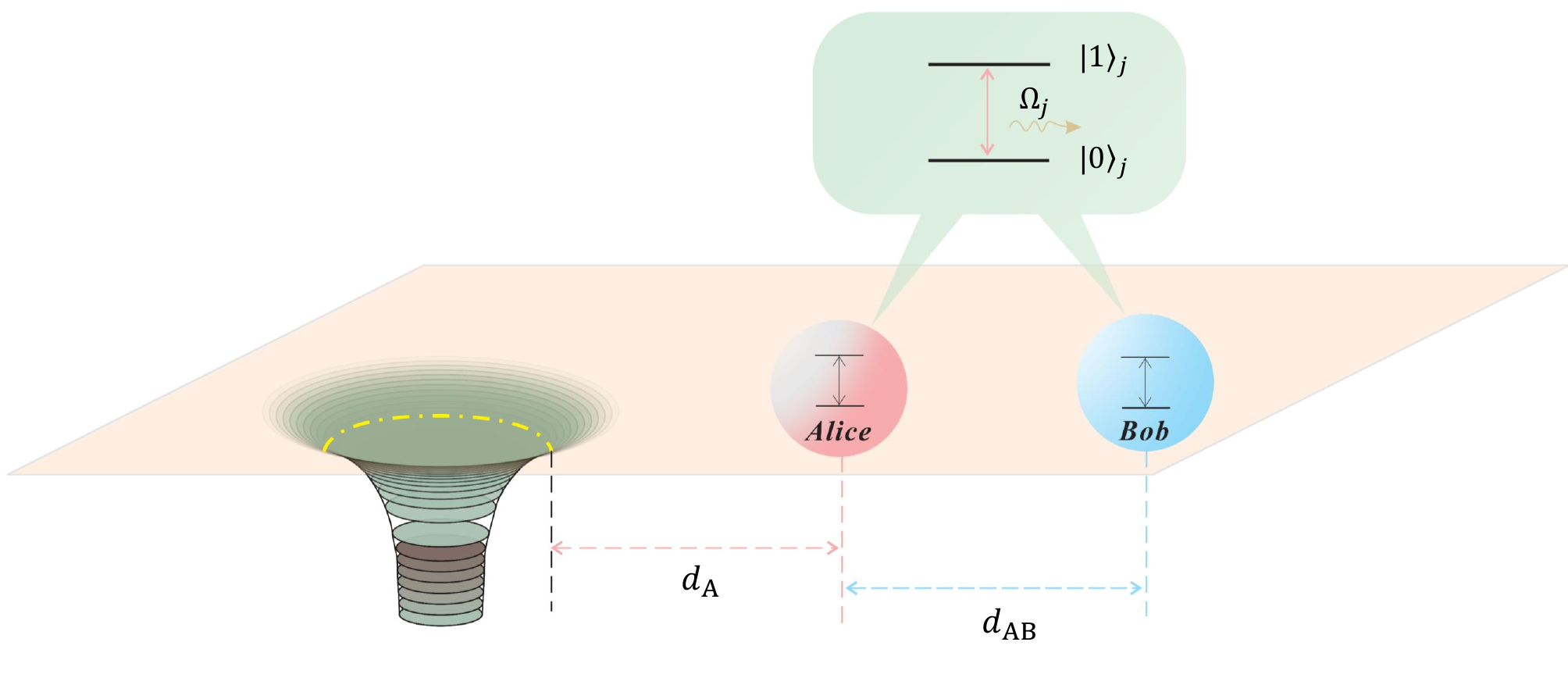} 
\caption{Schematic illustration of the UDW detectors arrangement in the correlations harvesting setup. 
Static detectors Alice and Bob are placed on the same side of the Lorentz-violating black hole, both initially uncorrelated in their ground states with an energy gap $ \Omega_j $.}
\label{fig1}
\end{figure*}

\section{Unruh-DeWitt detectors in Lorentz-violating spacetime}\label{sec3}
In this section, we provide a brief introduction to the UDW detector, which is a two-level quantum system with an energy gap $ \Omega $ between its ground and excited states. 
It interacts locally with the quantum scalar field along the detector's trajectory.
Let us now consider two point-like UDW detectors, which we label Alice and Bob, respectively.

First, we investigate how correlations between two UDW detectors depend on their proper separation. 
In particular, we place detector Alice closer to the Lorentz-violating BTZ-like black hole event horizon at $ r_h $, and detector Bob further out, ensuring $ r_\text{B}>r_\text{A}>r_h $.
Fig. \ref{fig1} illustrates this setup schematically: the pink surface represents the hypersurface on which the BTZ-like black hole is modeled and described by Eq. \meq{ds2}, and the yellow dashed circle denotes the location of the event horizon.
Each detector $ j\in\left\{ \text{A},\text{B} \right\} $ has a ground state $ \ket{0}_j $ and an excited state $ \ket{1}_j $, separated by an energy gap $ \Omega_j $. 
We define the proper distance between the detectors as
\begin{equation}
d_\text{AB}:=d\left(r_\text{A},r_\text{B}\right),
\end{equation}
which we hold fixed during our analysis. 
In the Lorentz-violating BTZ-like spacetime, the proper distance between the points $ (t, r_1, \varphi) $ and $ (t, r_2, \varphi) $ (with $ r_2>r_1>r_h $) is
\begin{equation}
d(r_1,r_2)=\int_{r_1}^{r_2}\!\sqrt{g_{\mu\nu}dx^\mu dx^\nu}=\ell\sqrt{1+\alpha} \ln\left[ \tfrac{r_2+\sqrt{r_2^2-r_h^2}}{r_1+\sqrt{r^2_1-r^2_h}} \right],
\end{equation}
where $ g_{\mu\nu} $ is the Lorentz-violating BTZ-like metric.
Similarly, the proper distances from the horizon to Alice and Bob can be written concisely as 
\begin{equation}
d_j:=d(r_h, r_j),\quad\text{for}\quad j\in\left\{\text{A},\text{B} \right\}.
\end{equation}

Furthermore, the Hawking temperature $ T_{H} $ of a static Lorentz-violating BTZ-like black hole is given by
\begin{align}
T_H=\frac{r_h}{2\pi\ell^2\sqrt{1+\alpha} },
\end{align}
and is related to the Lorentz-violating parameter $ \alpha $, decreasing as $ \alpha $ increases.
The local temperature $ T_j $ at the radial position $ r=r_j $ is defined as
\begin{align}\label{locT}
T_j=\frac{T_{H}}{\gamma_j},
\end{align}
which is recognized as the Kubo–Martin–Schwinger (KMS) temperature, where the redshift factor $ \gamma_j $ is given by
\begin{align}
\gamma_j=\sqrt{-g_{tt}}=\frac{1}{\ell}\sqrt{r^2_j-r^2_h},
\end{align}
subject to the condition $ r_j \geq r_h $.
Given $ r_\text{B}>r_\text{A}>r_h $, the redshift factors for detectors Alice and Bob can be expressed in terms of the proper distances as
\begin{equation}
\begin{aligned}
\gamma_\text{A}=\frac{r_h}{\ell}\sinh\frac{d_\text{A}}{\ell\sqrt{1+\alpha}},&&\gamma_\text{B}=\frac{r_h}{\ell}\sinh\frac{d_\text{AB}+d_{\text{A}}}{\ell\sqrt{1+\alpha}}.
\end{aligned}
\end{equation}

Subsequently, we consider the interaction between the detectors and the scalar fields.
Here, each UDW detector has its own proper time $ \tau_j $ (with $ j \in \{A, B\} $) and interacts with the local quantum field $ \hat{\phi}(\mathbf{x}) $ via the following interaction Hamiltonian in the interaction picture:
\begin{equation}\label{HHHH}
\hat{H}^{\tau_j}_j(\tau_j)=\lambda_j\Upsilon_j(\tau_j)\hat{\mu}_j(\tau_j)\otimes\hat{\phi}\left[\textbf{x}_j(\tau_j)\right],
\quad j\in\left\{ \text{A},\text{B} \right\},
\end{equation}
where the operator $ \hat{\mu}_j $, known as the monopole moment, describes the dynamics of a detector and is defined as:
\begin{equation}
\hat{\mu}_j(\tau_j)=\ket{1}_j\bra{0}_je^{i\Omega_j\tau_j}+\ket{0}_j\bra{1}_je^{-i\Omega_j\tau_j},
\end{equation}
and the switching function $ \Upsilon_j(\tau_j) $ is given by:
\begin{equation}
\Upsilon_j(\tau_j)=\exp\left[\tfrac{(\tau_j-\tau_{0,j})^2}{\sigma^2_j} \right],
\end{equation}
which governs the interaction duration for detector $ j $, with the parameter $ \tau_{0,j} $ denoting the peak of the switching function \cite{Gallock-Yoshimura:2021yok}, and the symbol $ \lambda_j $ represents the coupling strength.
Here, for simplicity, we assume that both detectors have the same coupling strength $ \lambda $, energy gap $ \Omega $, and switching duration $ \sigma_j = \sigma $ in their respective proper frames.
Note that the field operator $ \hat{\phi}[\mathbf{x}_j(\tau_j)] $ is the pullback of the quantum field along the trajectory of detector $ j $, and the superscript on the Hamiltonian $ \hat{H}^{\tau_j}_j(\tau_j) $ indicates that it generates time translations with respect to the detector's proper time $ \tau_j $.


Next, we express the total interaction Hamiltonian as the generator of time translation with respect to the common time $ t $ in the Lorentz-violating BTZ-like spacetime:
\begin{equation}
\hat{H}^t_{I}(t)=\frac{d\tau_\text{A}}{dt}\hat{H}^{\tau_\text{A}}_\text{A}\left[\tau_\text{A}(t)\right]+
\frac{d\tau_\text{B}}{dt}\hat{H}^{\tau_\text{B}}_\text{B}\left[\tau_\text{B}(t)\right],
\end{equation}
where we have used the time-reparametrization property \cite{Martin-Martinez:2020pss}. 
The time evolution is given by the unitary 
\begin{equation}
\hat{U}_I=\mathcal{T} \exp\left[ -i \int_{\mathbb{R}}dt \hat{H}^t_I(t) \right].
\end{equation}
where $ \mathcal{T} $ is a time-ordering symbol. 
Given that the coupling strength $ \lambda $ is small, we can expand the time evolution operator $ \hat{U}_I $ using a Dyson series,
\begin{equation}
\hat{U}_I=\mathbb{I}-i\int_\mathbb{R}dt\hat{H}^t_I(t)-\int_\mathbb{R}dt\int_{-\infty}^{t}dt'\hat{H}^t_I(t)\hat{H}^{t'}_I(t')
+\mathcal{O}(\lambda^3).
\end{equation}
Then, we assume that the detectors and the field are initially in their ground states and uncorrelated.
Consider a total system initialized in the state 
\begin{equation}
\rho_0=\ket{0}_\text{A}\bra{0}_\text{A}\otimes\ket{0}_\text{B}\bra{0}_\text{B}\otimes\ket{0}\bra{0},
\end{equation}
where $ \ket{0} $ is the field's vacuum state. 
After the interaction governed by $ \hat{U}_I $, the final total density matrix becomes
\begin{equation}
\rho_\text{tot}=\hat{U}_I~\rho_0~\hat{U}^{\dagger}_I=\rho_0+\sum_{i+j=1}^{2}\rho^{(i,j)}+\mathcal{O}(\lambda^4),
\end{equation}
where $ \rho^{(i,j)}=\hat{U}^{(i)}\rho_0\hat{U}^{(j)\dagger} $ and we used the fact that all the odd-power terms of $ \lambda $ vanish \cite{Pozas-Kerstjens:2015gta}.
Tracing out the field yields the detectors' reduced density matrix  $ \rho_\text{AB}=\text{Tr}_\phi\left[\rho_\text{tot} \right] $, which in the basis $ \left\{\ket{0}_\text{A}\ket{0}_\text{B},\ket{0}_\text{A}\ket{1}_\text{B},\ket{1}_\text{A}\ket{0}_\text{B},\ket{1}_\text{A}\ket{1}_\text{B} \right\}$, is known to take \cite{Martin-Martinez:2015qwa}
\begin{equation}\label{rho}
\rho_\text{AB} =
\begin{bmatrix}
1 - \mathcal{L}_\text{AA} - \mathcal{L}_\text{BB} & 0 & 0 & \mathcal{M}^* \\
0 & \mathcal{L}_\text{BB} & \mathcal{L}^*_\text{AB} & 0 \\
0 & \mathcal{L}_\text{AB} & \mathcal{L}_\text{AA} & 0 \\
\mathcal{M} & 0 & 0 & 0
\end{bmatrix}
+ \mathcal{O}(\lambda^4).
\end{equation}
The matrix elements are given by
\begin{equation}
\begin{aligned}\label{eq45}
\mathcal{L}_{ij}=&\lambda^2\int_{\mathbb{R}}d\tau_i\int_\mathbb{R}d\tau'_j \Upsilon_i(\tau_i)\Upsilon_j(\tau'_j)e^{-i\Omega(\tau_i-\tau'_j)}
\\&\times W^\text{Lv}_\text{BTZ}\left[\mathbf{x}_i(\tau_i),\mathbf{x}_j(\tau'_j) \right],
\end{aligned}
\end{equation}
\begin{equation}
\begin{aligned}\label{eq46}
\mathcal{M}=&-\lambda^2\int_{\mathbb{R}}d\tau_\text{A}\int_{\mathbb{R}}d\tau_\text{B}\Upsilon_\text{A}(\tau_\text{A})
\Upsilon_\text{B}(\tau_\text{B})e^{i\Omega(\tau_\text{A}+\tau_\text{B})}\\
&\times\bigg\{ \Theta\left[ t(\tau_\text{A})-t(\tau_\text{B}) \right]
W^\text{Lv}_\text{BTZ}\left[ \mathbf{x}_\text{A}(\tau_\text{A}), \mathbf{x}_\text{B}(\tau_\text{B}) \right]\\
&+\Theta\left[ t(\tau_\text{B})-t(\tau_\text{A}) \right]
W^\text{Lv}_\text{BTZ}\left[ \mathbf{x}_\text{B}(\tau_\text{B}), \mathbf{x}_\text{A}(\tau_\text{A}) \right] \bigg\},
\end{aligned}
\end{equation}
where $ \Theta(t) $ is the Heaviside step funtion.
The off-diagonal elements $ \mathcal{M} $ and $ \mathcal{L}_\text{AB} $ correspond to the nonlocal terms that depend on both trajectories, with $ \mathcal{M} $ responsible for entangling the two detectors and $ \mathcal{L}_\text{AB} $ used for calculating the mutual information.

It is important to note that, in the Lorentz-violating BTZ-like black hole spacetime, the quantities $ \mathcal{L}_\text{AB} $, $ \mathcal{L}_\text{AA} $, $ \mathcal{L}_\text{BB} $, and $ \mathcal{M} $ all depend on the Lorentz-violating parameter $ \alpha $.
We focus on static detectors along the same axis at the black hole's center $ (\Delta\phi=0) $, and assume both detectors switch on and off simultaneously. Under these conditions, the expression for $ \mathcal{L}_\text{AB} $ is given by
\begin{widetext}
\begin{equation}\label{LAB}
\begin{aligned}
\mathcal{L}_\text{AB}=&\lambda^2\int_{\mathbb{R}}d\tau_\text{A}\int_{\mathbb{R}}d\tau_\text{B}~e^{-\tau_\text{A}^2/2\sigma^2}e^{-\tau_\text{B}^2/2\sigma^2} e^{-i\Omega(\tau_\text{A}-\tau_\text{B})}
W^\text{Lv}_\text{BTZ}\left[\textbf{x}_\text{A}(\tau_\text{A}),\textbf{x}_\text{B}(\tau_\text{B})\right]\\
=&\frac{\lambda^2}{4\pi\ell\sqrt{2(1+\alpha)}}\sum_{n=-\infty}^{\infty}\int_\mathbb{R}d\tau_\text{A}
\int_\mathbb{R}d\tau_\text{B}~e^{-\tau_\text{A}^2/2\sigma^2}e^{-\tau_\text{B}^2/2\sigma^2}e^{-i\Omega\left(\tau_\text{A}-\tau_\text{B}\right)}
\left[ \frac{1}{\rho_1^{-}\left(\textbf{x}_\text{A},\Gamma^n\textbf{x}_\text{B}\right)}-\frac{\zeta}{\rho_1^{+}\left(\textbf{x}_\text{A},\Gamma^n\textbf{x}_\text{B}\right)} \right]\\
=&\frac{\lambda^2\gamma_\text{A}\gamma_\text{B}}{4\pi\ell\sqrt{2(1+\alpha)}}
\sum_{n=-\infty}^{\infty}\int_\mathbb{R}dt_\text{A}\int_\mathbb{R}dt_\text{B}
~e^{-\gamma_\text{A}^2t_\text{A}^2/2\sigma^2}e^{-\gamma_\text{B}^2t_\text{B}^2/2\sigma^2}e^{-i\Omega\left(\gamma_\text{A}t_\text{A}-\gamma_\text{B}t_\text{B}\right)}\left[\frac{1}{\rho_2^-(t_\text{A},t_\text{B})}-\frac{\zeta}{\rho_2^+(t_\text{A},t_\text{B})} \right]\\
=&\frac{\lambda^2\gamma_\text{A}\gamma_\text{B}}{8\pi\ell\sqrt{2(1+\alpha)}}
\sum_{n=-\infty}^{\infty}\int_{\mathbb{R}}du~e^{-(\gamma_\text{A}^2+\gamma_\text{B}^2)\frac{u^2}{8\sigma^2}}
e^{-i\Omega(\gamma_\text{A}+\gamma_\text{B})\frac{u}{2}}\left[\frac{1}{\rho_3^-(u)}-\frac{\zeta}{\rho_3^+(u)}\right]
\int_\mathbb{R}dv~e^{-(\gamma_\text{A}^2+\gamma_\text{B}^2)\frac{v^2}{8\sigma^2}}e^{-(\gamma_\text{A}^2-\gamma_\text{B}^2)\frac{uv}{4\sigma^2}}e^{-i\Omega(\gamma_\text{A}-\gamma_\text{B})\frac{v}{2}}\\
=&\frac{\lambda^2\sigma\gamma_\text{A}\gamma_\text{B}}{4\ell\sqrt{\pi(1+\alpha)}}
\tfrac{1}{\sqrt{(\gamma^2_\text{A}+\gamma_\text{B}^2)}}
\exp\left[-\tfrac{\Omega^2\sigma^2(\gamma_\text{A}-\gamma_\text{B})^2}{2(\gamma_\text{A}^2+\gamma_\text{B}^2)}\right]
\sum_{n=-\infty}^{\infty}\int_\mathbb{R}
du~\exp\left[{-\tfrac{\gamma^2_\text{A}\gamma^2_\text{B}u^2}{2\sigma^2(\gamma^2_\text{A}+\gamma^2_\text{B})}}\right]
\exp\left[{-\tfrac{i\gamma_\text{A}\gamma_\text{B}(\gamma_\text{A}+\gamma_\text{B})u}{\gamma^2_\text{A}+\gamma^2_\text{B}} }\right]
\left[\frac{1}{\rho_3^-(u)}-\frac{\zeta}{\rho_3^+(u)}\right]\\
=&2K\sum_{n=-\infty}^{\infty}\text{Re}\int^{\infty}_{0}dx e^{-ax^2}e^{-i\beta x}
\left[\left(\cosh\chi^{-}_\text{AB,n}-\cosh x\right)^{-1/2}-\zeta\left(\cosh\chi^{+}_\text{AB,n}-\cosh x\right)^{-1/2} \right],
\end{aligned}
\end{equation}
\end{widetext}
where $ a $, $ \beta $ and $ \chi^\pm_\text{AB,n} $ are defined by
\begin{equation}
\begin{aligned}
&a:=\frac{\gamma^2_\text{A}\gamma^2_\text{B}}{2\sigma^2(\gamma^2_\text{A}+\gamma^2_\text{B})}\frac{\ell^4(1+\alpha)}{r_h^2},
\end{aligned}
\end{equation}
\begin{equation}
\begin{aligned}
&\beta:=\frac{\gamma_\text{A}\gamma_\text{B}(\gamma_\text{A}+\gamma_\text{B})}{\gamma^2_\text{A}+\gamma^2_\text{B}}\frac{\ell^2\sqrt{1+\alpha}}{r_h}\Omega,
\end{aligned}
\end{equation}
\begin{equation}
\begin{aligned}
&K:=\frac{\lambda^2\sigma}{4}\sqrt{\tfrac{\gamma_\text{A}\gamma_\text{B}}{\pi(\gamma^2_\text{A}+\gamma^2_\text{B})}}
\exp\left[-\tfrac{\Omega^2\sigma^2(\gamma_\text{A}-\gamma_\text{B})^2}{2(\gamma^2_\text{A}+\gamma^2_\text{B})} \right],
\end{aligned}
\end{equation}
\begin{equation}
\begin{aligned}
\chi^{\pm}_\text{AB,n}:=\text{arccosh}\left[\frac{r_h^2}{\ell^2\gamma_\text{A}\gamma_\text{B}}
\left(\frac{r_\text{A}r_\text{B}}{r^2_h}\cosh\left[\tfrac{2\pi n r_h}{\ell\sqrt{1+\alpha}}\right]\pm1\right) \right],
\end{aligned}
\end{equation}
and $ \rho^\pm_{1,2,3} $ are
\begin{equation}
\begin{aligned}
\rho_1^{\pm}:=&\sqrt{\sigma_n\left(\textbf{x}_\text{A},\Gamma^n\textbf{x}_\text{B}\right)+1\pm1} \Big|_{(r=r_\text{A},r'=r_\text{B})}\\
=&\sqrt{\tfrac{r_\text{A}r_\text{B}}{r^2_h}\cosh\left[\tfrac{2\pi n r_h}{\ell\sqrt{1+\alpha}}\right]\pm1
-\tfrac{\ell^2\gamma_\text{A}\gamma_\text{B}}{r_h^2}
\cosh\left(\tfrac{r_h \Delta t}{\ell^2\sqrt{1+\alpha}}\right)},\\
\rho_2^{\pm}=&\frac{\ell\sqrt{\gamma_\text{A}\gamma_\text{B}}}{r_h}
\left[\cosh\chi^{\pm}_\text{AB,n}-\cosh\left(\tfrac{r_h(t_\text{A}-t_\text{B})}{\ell^2\sqrt{1+\alpha}}\right)\right]^{1/2},\\
\rho_3^{\pm}=&\frac{\ell\sqrt{\gamma_\text{A}\gamma_\text{B}}}{r_h}
\left[\cosh\chi^{\pm}_\text{AB,n}-\cosh\left(\tfrac{r_h u}{\ell^2\sqrt{1+\alpha}}\right)\right]^{1/2}.
\end{aligned}
\end{equation}
The intermediate steps involve changing variables $ \tau_j\rightarrow t_j $ via $ t_j:=\tau_j/\gamma_j $ and then using the transformations $ u:=t_\text{A}-t_\text{B} $ and $ v:=t_\text{A}+t_\text{B} $ to reduce the double integral to a single integral.
By treating $ dt_\text{A} $ and $ dt_\text{B} $ as 1-forms, one finds 
\begin{equation}
dt_\text{A}dt_\text{B}=\frac{1}{2}dudv,
\end{equation}
reflecting the Jacobian factor of $ 1/2 $. 
After completing the above coordinate transformation, the integral can be evaluated over the variable $v$.
Finally, defining $x := \frac{r_h}{\ell^2\sqrt{1+\alpha}}u$ enables the final form of $ \mathcal{L}_\text{AB} $ for numerical evaluation.

Similarly, one can derive the numerical evaluation form of $ \mathcal{L}_\text{AA} $, $ \mathcal{L}_\text{BB} $, and $ \mathcal{M} $ under Lorentz violation. For instance, considering the identical expressions of $ \mathcal{L}_\text{AA} $ and $ \mathcal{L}_\text{BB} $, we use $ \mathcal{L}_\text{DD} $ to represent them, which is given by
\begin{equation}\label{LDD}
\begin{aligned}
\!\mathcal{L}_\text{DD}=&-\frac{\zeta\lambda^2\sigma}{2\sqrt{2\pi}}\text{Re}\int_0^{\infty}dx\frac{e^{-a_\text{D}x^2}e^{-i\beta_\text{D} x}}{\sqrt{\cosh \chi^+_\text{D,0}-\cosh x}}\\
&+\frac{\lambda^2\sigma^2}{2}\int_\mathbb{R} dx\frac{e^{-\sigma^2(x-\Omega)^2}}{e^{x/T_\text{D}}+1}
+\frac{\lambda^2\sigma}{\sqrt{2\pi}}\sum_{n=1}^{\infty}\text{Re}\int_{0}^{\infty}dx \\
&\times e^{-a_\text{D}x^2}e^{-i\beta_\text{D}x}\left(\tfrac{1}{\sqrt{\cosh\chi^-_\text{D,n}-\cosh x}}-\tfrac{\zeta}{\sqrt{\cosh\chi^-_\text{D,n}-\cosh x}}\right),
\end{aligned}
\end{equation}
where $ T_\text{D}=r_h/\left(2\pi\ell^2\gamma_\text{D}\sqrt{1+\alpha} \right)$ is the local temperature at $ r=r_\text{D} $ and 
\begin{equation}
\begin{aligned}
&a_\text{D}:=\frac{\ell^4\gamma^2_\text{D}(1+\alpha)}{4\sigma^2 r_h^2}, \quad
\beta_\text{D}:=\frac{\sqrt{1+\alpha}\ell^2\gamma_\text{D}\Omega}{r_h},\\
&\chi^{\pm}_\text{D,n}:=\text{arccosh}\left[\frac{r_h^2}{\ell^2\gamma_\text{D}^2}
\left(\frac{r_\text{D}^2}{r^2_h}\cosh\left[\tfrac{2\pi n r_h}{\ell\sqrt{1+\alpha}}\right]\pm1\right) \right].
\end{aligned}
\end{equation}
The first two terms, corresponding to $ (n=0) $, resemble AdS–Rindler contributions in a Lorentz-violating spacetime, whereas the last term $ (n\neq0) $ is known as the BTZ-like term. 
Both the second and third integrals in \eqref{LDD} exhibit the same branch cut subtlety as $ \mathcal{L}_\text{AB} $, but can be handled in an analogous manner \cite{Bueley:2022ple}. 
When the two detectors are at the same position, i.e., $ d_\text{AB}=0 $, the numerical result of $ \mathcal{L}_\text{AB} $ will be identical to the above equation.
One also obtains the matrix element $ \mathcal{M} $  (used in the concurrence calculation) in a explicit form:
\begin{equation}
\begin{aligned}
\mathcal{M}=-\sum_{n=-\infty}^{\infty}&\left[K_\text{M}\int_{0}^{\infty}dx\frac{\exp({-a_\text{M}x^2})\cos(\beta_\text{M}x)}{\sqrt{\cosh\chi^-_\text{M,n}-\cosh x}}\right. \\
&\left.-\zeta K_\text{M}\int_{0}^{\infty}dx\frac{\exp({-a_\text{M}x^2})\cos(\beta_\text{M}x)}{\sqrt{\cosh\chi^+_\text{M,n}-\cosh x}} \right],
\end{aligned}\label{eq56}
\end{equation}
with corresponding definitions for
\begin{equation}
\begin{aligned}
&K_\text{M}:=\frac{2\sqrt{\pi}}{\pi}K,\quad \chi^{\pm}_\text{M,n}:=\chi^{\pm}_\text{AB,n},\quad a_\text{M}:=a,\\
&\beta_\text{M}:=\frac{\gamma_\text{A}\gamma_\text{B}(\gamma_\text{A}-\gamma_\text{B})}{\gamma^2_\text{A}+\gamma^2_\text{B}}\frac{\ell^2\sqrt{1+\alpha}}{r_h}\Omega.
\end{aligned}
\end{equation}
Here again, the Lorentz-violating parameter $ \alpha $ modifies the spacetime structure and thus influences the integrals that characterize each physical quantity. Each of these integrals $ (\mathcal{L}_\text{AB}, \mathcal{L}_\text{DD}, $ and $ \mathcal{M}) $ follows a similar procedure: apply coordinate transformations to simplify the double integrals and identify the relevant Jacobian factors. 
The Lorentz-violating parameter $ \alpha $ alters the effective Lorentz symmetry in the BTZ framework, resulting in modifications to the spacetime geometry and metric.
These changes affect the Hawking temperature and the Green's function terms, leading to numerical results that differ from the standard BTZ scenario.

\section{Results}\label{sec4}
To quantify the influence of Lorentz violation on the extraction of entanglement after the interaction of the space-like separated detectors with the Lorentz-violating vector field, we utilize concurrence as the entanglement measure, which, with the density matrix \meq{rho} is \cite{Martin-Martinez:2015qwa,Henderson:2018lcy}
\begin{equation}\label{eq58}
\mathcal{C}(\rho_\text{AB})=2\left[0, \left(\left|\mathcal{M}\right|-\sqrt{\mathcal{L}_\text{AA}\mathcal{L}_\text{BB}}\right) 
 \right]+\mathcal{O}(\lambda^4).
\end{equation}
Clearly, the concurrence is a competition between the correlation term $ \mathcal{M} $ and the detector's transition probabilities $ \mathcal{L}_\text{AA}$ and $\mathcal{L}_\text{BB} $.

The total correlations (including both classical correlations and quantum correlations) between detectors is described by mutual information, which is defined as
\begin{equation}
\begin{aligned}
\mathcal{I}(\rho_\text{AB})=&\mathcal{L}_{+}\ln \mathcal{L}_{+}+\mathcal{L}_{-}\ln \mathcal{L}_{-}\\
&-\mathcal{L}_\text{AA}\ln \mathcal{L}_\text{AA}
-\mathcal{L}_\text{BB}\ln \mathcal{L}_\text{BB}+\mathcal{O}(\lambda^4),
\end{aligned}\label{eq59}
\end{equation}
with
\begin{equation}
\mathcal{L}_{\pm}:=\frac{1}{2}\left[\mathcal{L}_\text{AA}+\mathcal{L}_\text{BB}\pm 
\sqrt{\left(\mathcal{L}_\text{AA}-\mathcal{L}_\text{BB}\right)^2+4\left|\mathcal{L}_\text{AB}\right|^2}  \right].
\end{equation}
Note that, unlike the concurrence $ \mathcal{C}(\rho_\text{AB}) $, the mutual information $ \mathcal{I}(\rho_\text{AB}) $ is determined by the transition probabilities $ \mathcal{L}_\text{AA}$, $\mathcal{L}_\text{BB} $ and the correlation term $ \mathcal{L}_\text{AB}$. 
And from the Eq. (\ref{eq59}) knows that if the correlation term $ \mathcal{L}_\text{AB}=0$, then $ \mathcal{I}(\rho_\text{AB})=0 $. 
Furthermore, mutual information persists even when the concurrence $ \mathcal{C}(\rho_\text{AB})=0 $, at which time the extracted correlations between detectors are either classical correlations or non-distillable entanglement. n particular, in this work, the correlation terms ($ (\mathcal{L}_\text{AB}, \mathcal{L}_\text{DD}, $ and $ \mathcal{M}) $) are involved in the convergence of the image summation. To ensure the accuracy of the numerical results, we performed a convergence test on the number of terms. The results show that when the number of image terms is set to 5, the correlation terms have converged to three decimal places, and the numerical results are stable. Adding further summation terms will not significantly alter the computational results, but will increase processing time and resource usage. Therefore, we set the number of image summation terms to 5 in the following calculations to improve computational efficiency while ensuring accuracy.

\subsection{Entanglement harvesting}
We now consider harvested entanglement in the case of the detectors subjected to the Lorentz violation when the proper distance between the detector Alice and horizon causes a change. 
The transition probabilities $ \mathcal{L}_\text{AA} $ and $ \mathcal{L}_\text{BB} $, as well as the matrix element $ \mathcal{M} $, may be derived numerically using Eqs. (\ref{eq45}) and (\ref{eq46}), after which the concurrence given by Eq. (\ref{eq58}) can be easily assessed the generation of entanglement between the detectors.

\begin{figure}[h]
\centering
\includegraphics[scale=0.41]{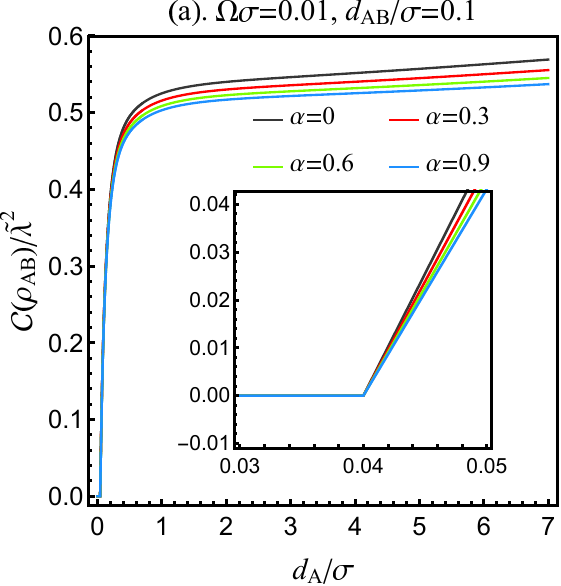}
\includegraphics[scale=0.42]{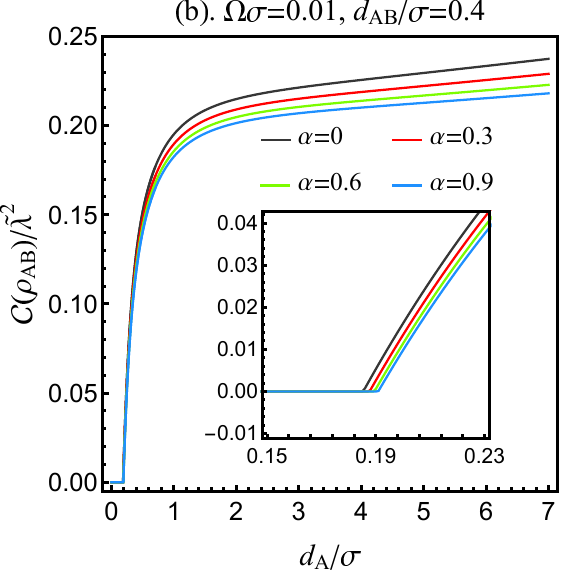}
\includegraphics[scale=0.41]{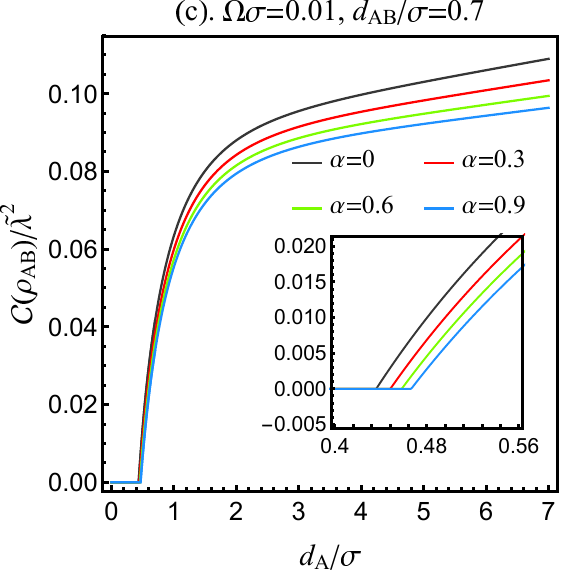}
\includegraphics[scale=0.42]{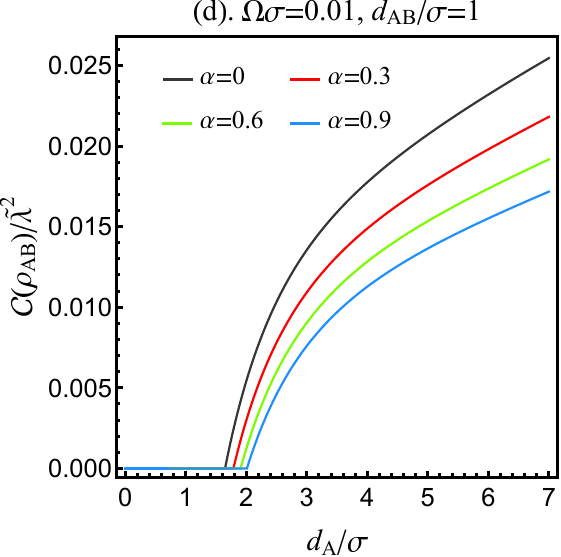}
\caption{ The concurrence $\mathcal{C}(\rho_{\mathrm{AB}})/\tilde{\lambda}^2$ between two detectors in the Lorentz-violating BTZ-like spacetime as a function of distance $d_{\mathrm{A}}$ from horizon is plotted for different values of $\alpha$ and $\tilde{\lambda}:=\lambda\sqrt{\sigma}$ is the dimensionless coupling strength. 
In all the legends, we set $\ell/\sigma=10$ and $ M = 0.01 $.}
\label{fig2}
\end{figure}\noindent	

In Fig. \ref{fig2}, the amount of obtained entanglement is plotted as a function of the appropriate distances of the detector Alice from the horizon for various Lorentz-violating parameters $\alpha$.
As expected, the closer detector Alice is to the horizon, the less entanglement can be harvested.
For fixed energy gaps of the detectors, we find that compared with the BTZ black holes uncoupled to Lorentz-violating vector fields, the presence of Lorentz violation suppresses the entanglement extraction between detectors in BTZ-like black holes.
Note that due to the intrinsic properties of black holes, including Hawking temperature and gravitational redshift, which inhibit entanglement harvesting \cite{Henderson:2017yuv}, there is a feasible range of entanglement that can be extracted between detectors, i.e., a critical proper distance $d_{\mathrm{death}}\left(r_{h},R_\text{A}\right)/\sigma$ exists, below which entanglement harvesting occurs in a ``sudden death", a range we refer to as the \textit{entanglement shadow}.
\begin{figure}[h]
\centering
\includegraphics[scale=0.42]{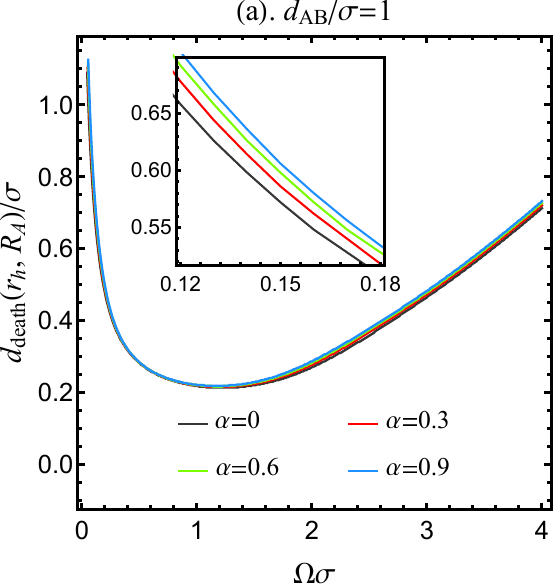}
\includegraphics[scale=0.42]{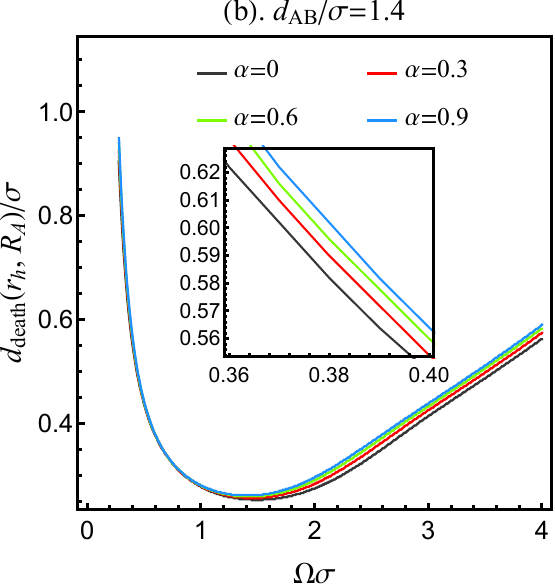}
\includegraphics[scale=0.43]{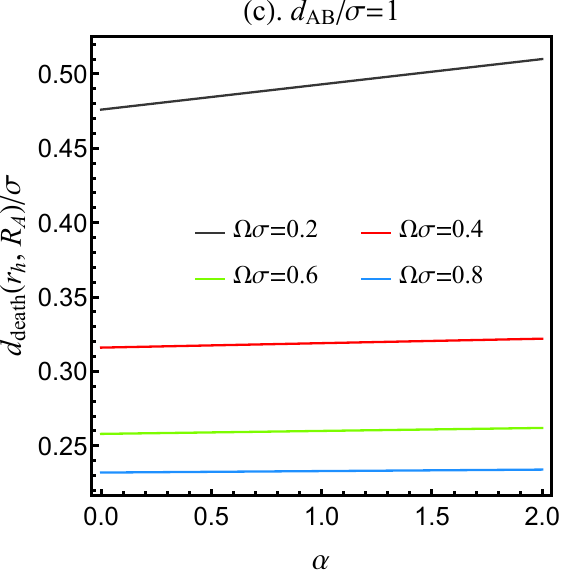}
\includegraphics[scale=0.42]{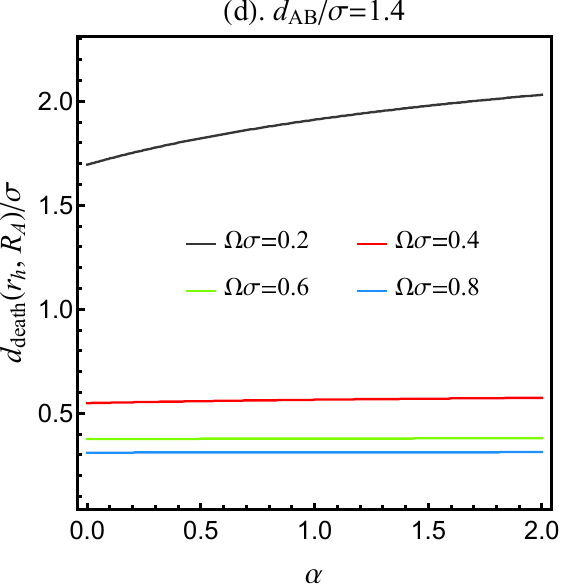}
\caption{(a), (b): The critical distance $d_{\mathrm{death}}\left(r_{h},R_\text{A}\right)/\sigma$ for entanglement harvesting as a function of energy gaps $\Omega\sigma $. 
(c), (d): The $d_{\mathrm{death}}\left(r_{h},R_\text{A}\right)/\sigma$ as a function of Lorentz-violating parameter $\alpha$ for various $\Omega\sigma $.}
\label{fig3}
\end{figure}\noindent
The figure shows that decreasing the inter-detectors separation can significantly reduce the range of entanglement shadow, which suggests that smaller distances between detectors usually have the advantage in harvested more entanglement compared to larger distances.
To gain a clear understanding of the crucial range of the ``sudden death" of entanglement, we show in Fig. \ref{fig3} how it changes with respect to the detectors' energy gaps and the Lorentz-violating parameter. 
It is easy to see that the influence of the energy gap on entanglement shadow is not monotonic [see Figs. \ref{fig3} a, b]. 
Therefore, determining the optimal energy gap facilitates the extraction of greater amounts of entanglement. 
Obviously, the critical range of entanglement shadow is an increasing function of the Lorentz violation, indicating that the presence of Lorentz violation broadens the scope of entanglement ``sudden death".

\begin{figure}[h]
\centering
\includegraphics[scale=0.41]{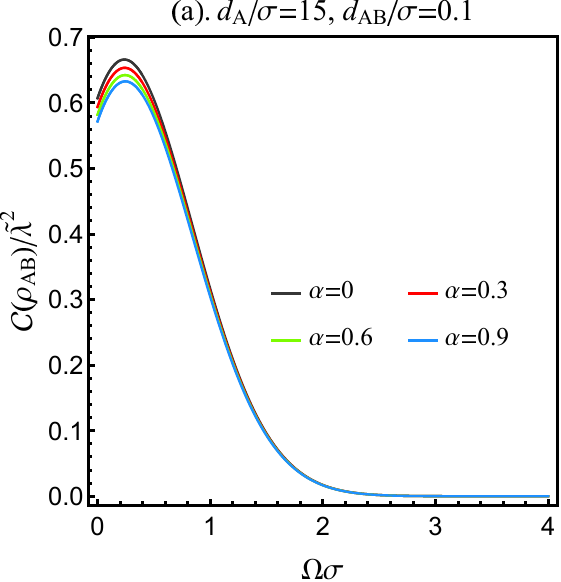}
\includegraphics[scale=0.42]{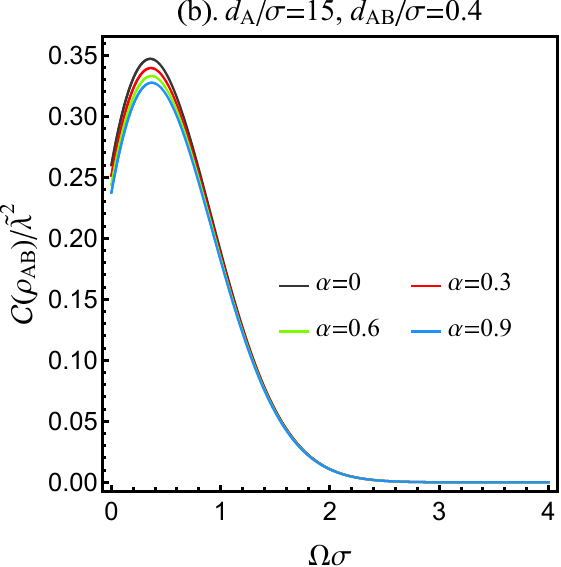}
\includegraphics[scale=0.41]{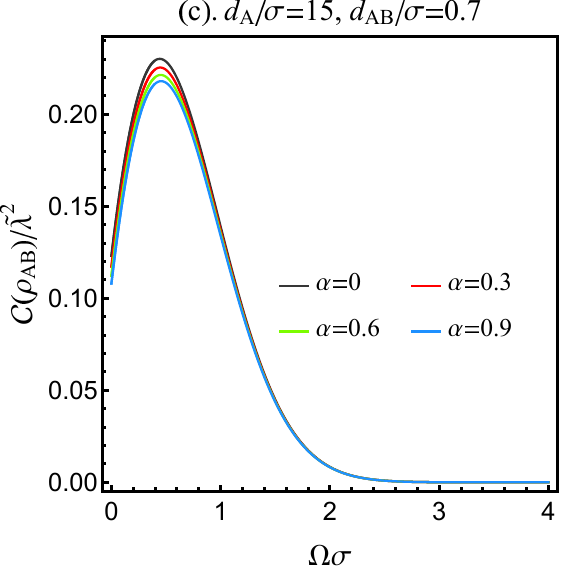}
\includegraphics[scale=0.42]{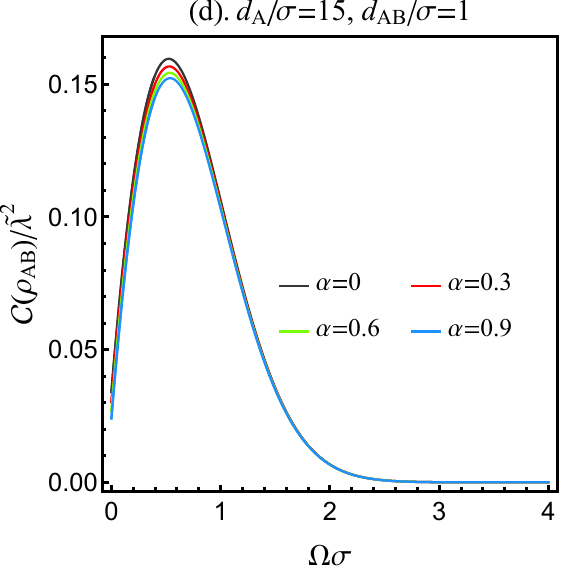}
\caption{ The concurrence is plotted as a function of detectors' energy gap $\Omega\sigma$ for various values of $\alpha$. 
Here, we set proper spacing between detector Alice and the horizon is $d_{\mathrm{A}}/\sigma$ = 15.}
\label{fig4}
\end{figure}

In Fig. \ref{fig4}, we display the effect of the detectors' energy gaps on harvested entanglement by plotting the concurrence as a function of $\Omega$ for different Lorentz-violating parameters. 
As indicated in the image, when detector Alice is at an appropriate distance from the horizon, there is always an optimal energy gap that allows harvesting the greatest amount of entanglement.
\noindent
And at the optimal detectors' energy gap, we find that Lorentzian violation has the most obvious effect on the amount of entanglement harvesting. 
\begin{figure}[h]
\centering
\includegraphics[scale=0.41]{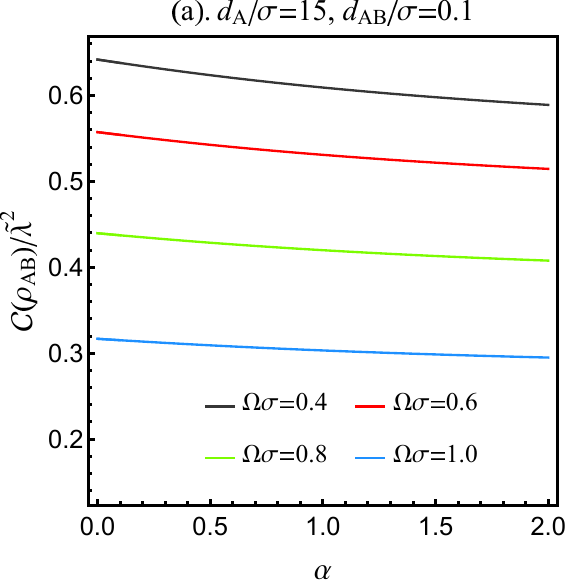}
\includegraphics[scale=0.42]{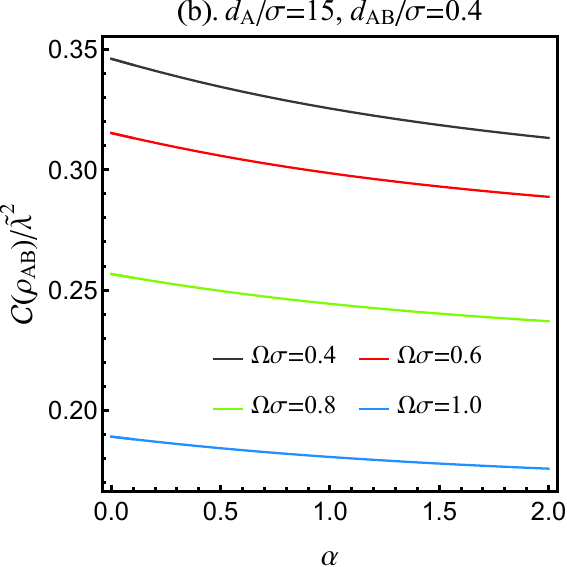}
\includegraphics[scale=0.42]{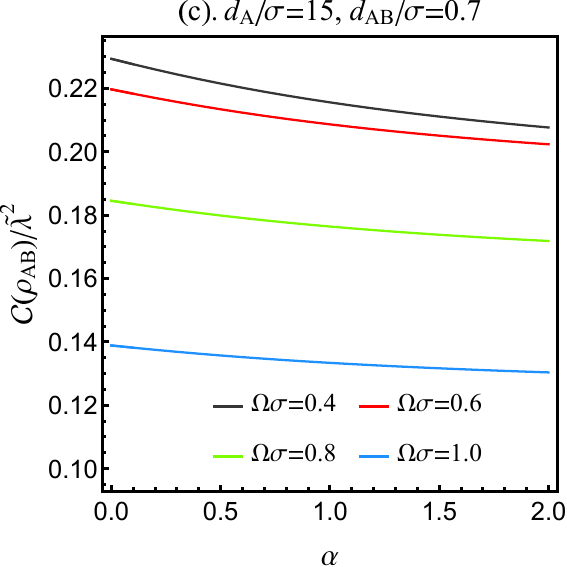}
\includegraphics[scale=0.42]{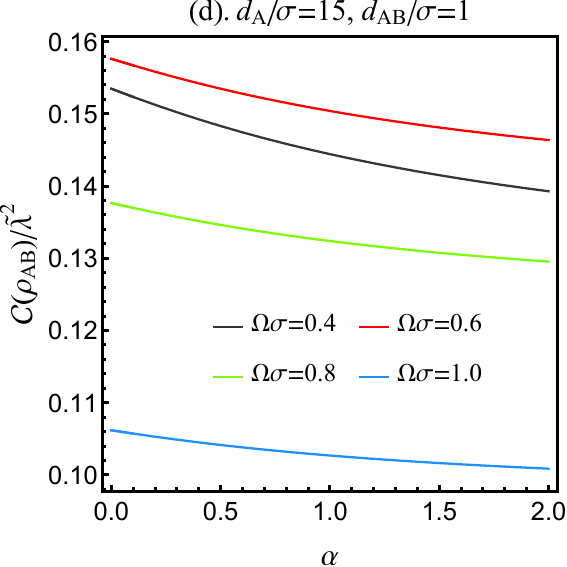}
\caption{ The concurrence $\mathcal{C}(\rho_{\mathrm{AB}})$ as a function of the Lorentz-violating parameter $\alpha$ with $d_{\mathrm{A}}/\sigma$ = 15 for various detectors' energy gaps $\Omega\sigma$.}
\label{fig5}
\end{figure}\noindent
The impact of Lorentz violation on concurrence is then significantly weakened when either reducing or increasing the energy gap, indicating that the energy gap of the detector influences the sensitivity of Lorentz violation to entanglement harvesting. 
Furthermore, as in the previous result of fixed energy gaps, the presence of Lorentz violation when fluctuations in the energy gap also inhibits the harvested entanglement.

To further investigate how the phenomenon of entanglement harvesting depends on Lorentz violation, we present the concurrence as a function of the Lorentz-violating parameter $\alpha$ in Fig. \ref{fig5}.
The figure shows that the amount of entanglement harvesting between the detectors is always a decreasing function as the Lorentz-violating parameter increases, regardless of the interdetector separation. 
This is consistent with the result of the suppression of entanglement harvesting due to Lorentz violation shown in the figure above.
Meanwhile, we can visualize the effect of the energy gap on the sensitivity of Lorentz violation more intuitively from the figure. 
Corresponding to Fig. \ref{fig4}, it shows that the sensitivity of Lorentz violation to entanglement extraction decreases after exceeding the optimal energy gap.

\subsection{Mutual information harvesting}

Mutual information quantifies the total amount of classical and quantum correlations, including entanglement. 
By analyzing mutual information and entanglement together, we can uncover key differences between the effects of Lorentz violation on classical and quantum correlations.
Now, we begin to numerically evaluate the mutual information harvesting between detectors as given in Eq. (\ref{eq59}). 
\begin{figure}[h]
\centering
\includegraphics[scale=0.41]{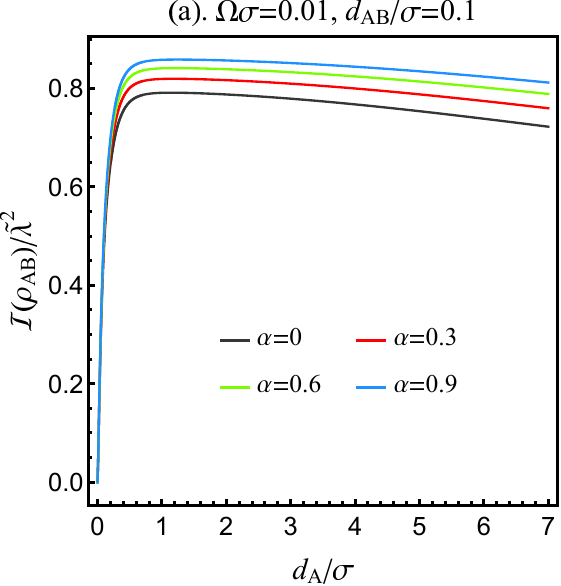}
\includegraphics[scale=0.42]{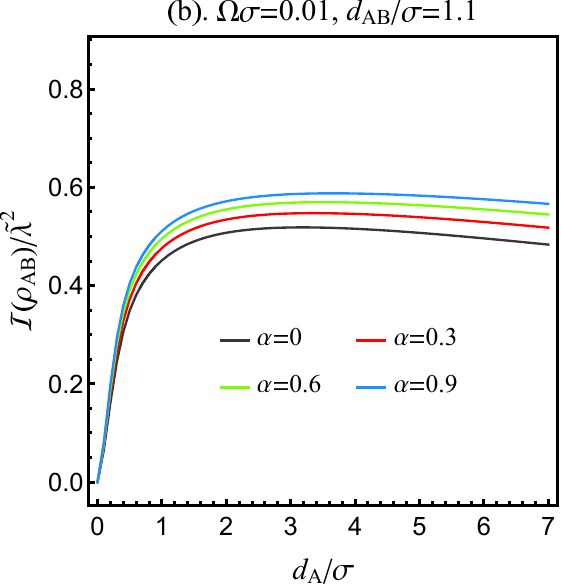}
\includegraphics[scale=0.42]{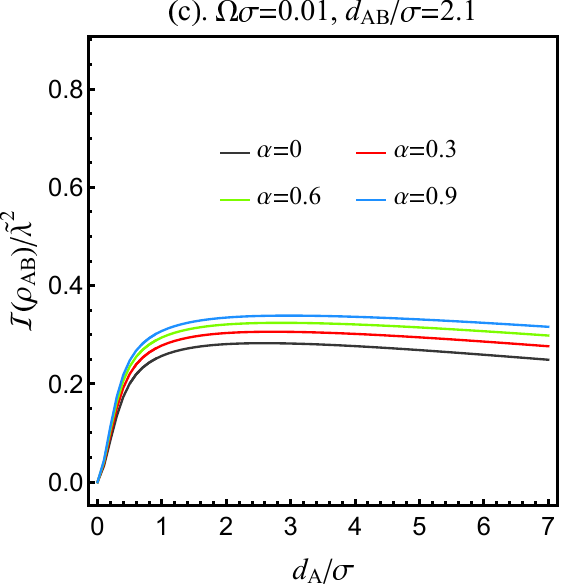}
\includegraphics[scale=0.42]{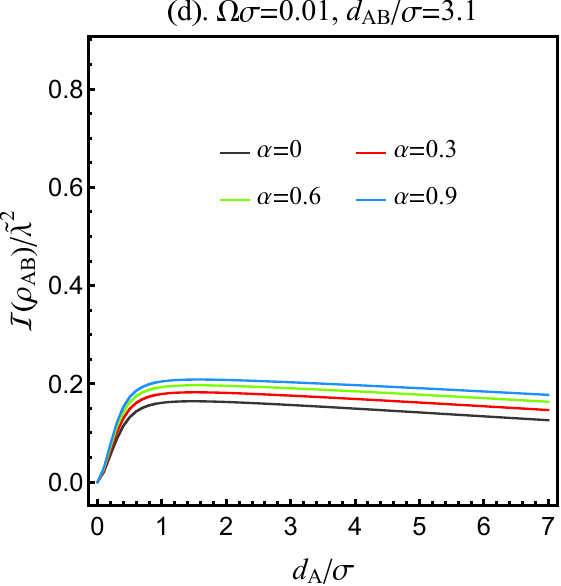}
\caption{ The plots of mutual information $\mathcal{I} (\rho_{\mathrm{AB}})$ versus the proper distance detector Alice is from the horizon $d_{\mathrm{A}}/\sigma$ for various Lorentz-violating parameters $\alpha$.}
\label{fig6}
\end{figure}\noindent

\begin{figure}[h]
\centering
\includegraphics[scale=0.41]{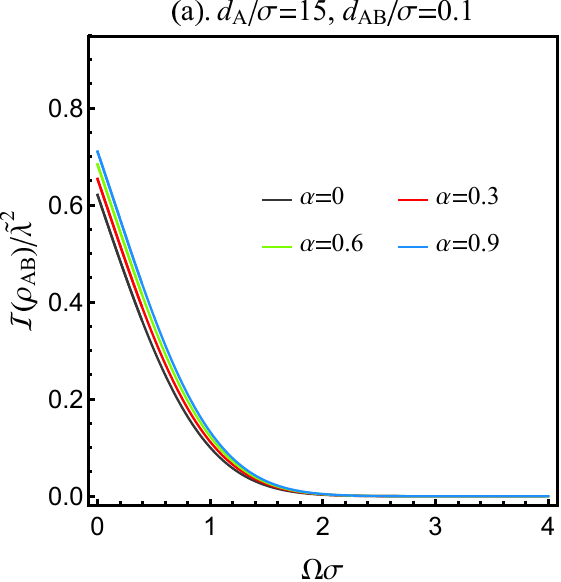}
\includegraphics[scale=0.42]{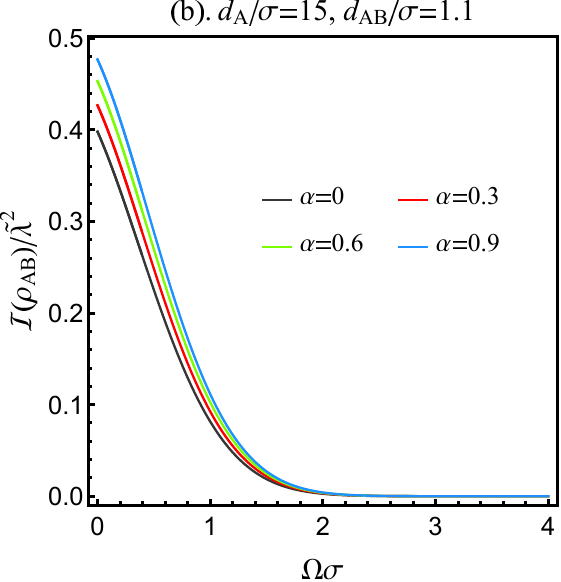}
\includegraphics[scale=0.42]{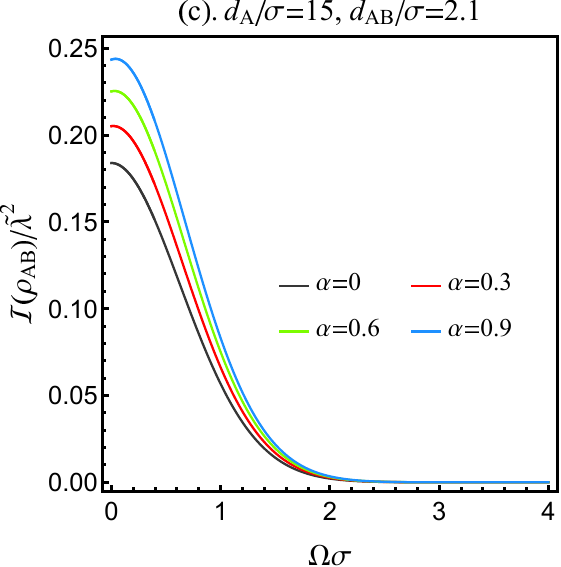}
\includegraphics[scale=0.42]{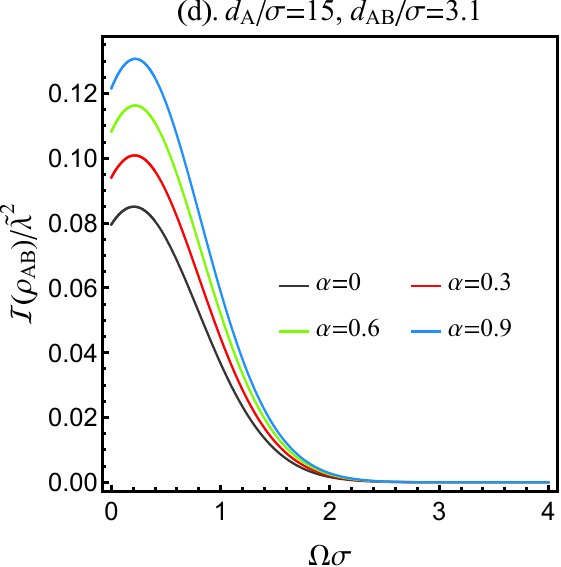}
\caption{ The mutual information is plotted as the function of the energy gap $\Omega\sigma$ for different Lorentz-violating parameters $\alpha$ with the fixed distance between detector Alice and the horizon is $d_{\mathrm{A}}/\sigma=15$.}
\label{fig7}
\end{figure}\noindent

Figs. \ref{fig6} and \ref{fig7} show the amount of mutual information harvesting as a function of detector-to-horizon distance and energy gaps of the detectors for various Lorentz-violating parameters. 
However, unlike the trend of monotonic increasing in entanglement harvesting with increasing detector distance from the horizon, at a fixed energy gap, the amount of mutual information harvesting will rise sharply to a peak and then gradually decline. 
Furthermore, it is clear that the inter-detector separation has a significant impact on the amount of entanglement harvesting, as larger inter-detector separations can decrease the amount of extracted entanglement. 
Fig. \ref{fig7} demonstrates that at smaller inter-detector separations, there is no optimal energy gap that is the same as the one for harvesting entanglement [see Figs. \ref{fig7} a, b]. 
Notably, we discover that in the context of Lorentz-violating BTZ-like black hole, the presence of Lorentz violation significantly enhances the amount of mutual information harvesting compared to the standard BTZ black hole, which is contrary to its effect on entanglement harvesting. 

\begin{figure}[h]
\centering
\includegraphics[scale=0.42]{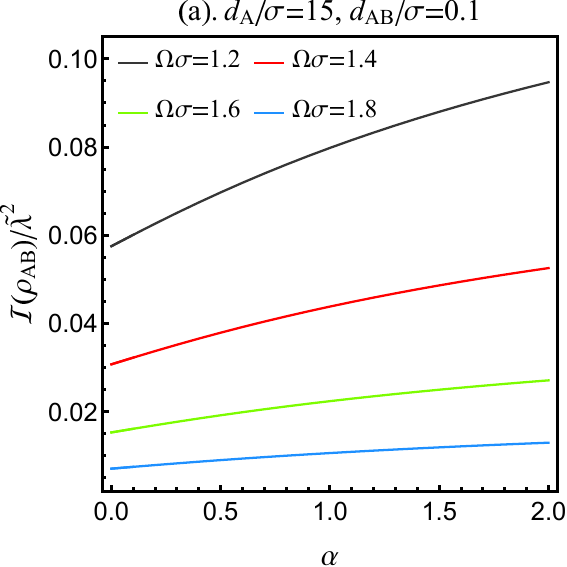}
\includegraphics[scale=0.42]{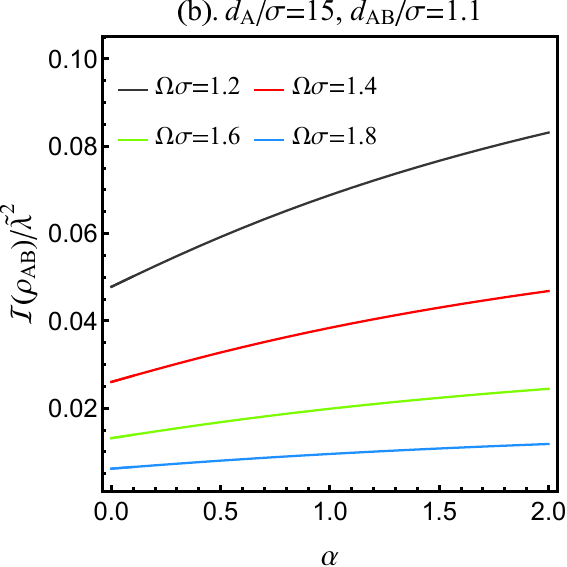}
\includegraphics[scale=0.42]{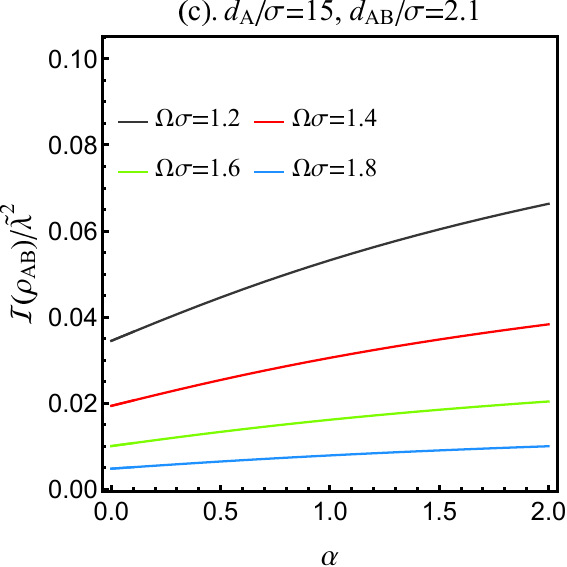}
\includegraphics[scale=0.42]{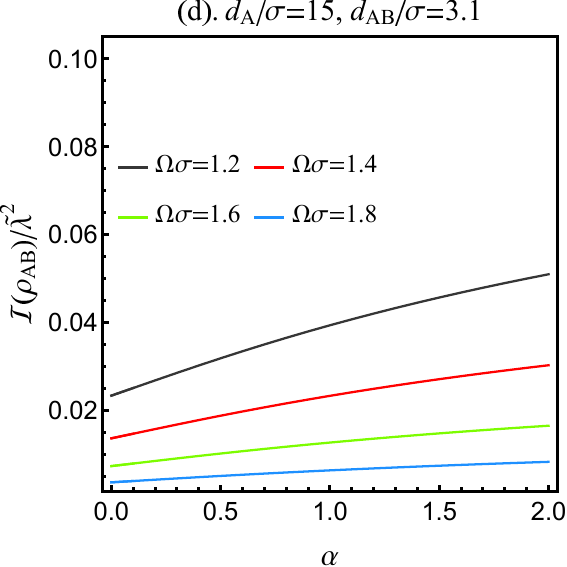}
\caption{ Mutual information $ \mathcal{I} $ is plotted as the function of Lorentz-violating parameter $\alpha$ for various detectors' energy gaps with $d_{\mathrm{A}}/\sigma=15$.
}\label{fig8}
\end{figure}\noindent

To further study the influence of Lorentz violation on mutual information harvesting, we plot mutual information versus the Lorentz-violating parameter in Fig. \ref{fig8}. 
From the figure, we can intuitively see that for the non-fixed detectors' energy gaps, the amount of harvested mutual information is always an increasing function of Lorentz violation, irrespective of the inter-detector separation, indicating that Lorentz violation enhances the amount of extracted mutual information, which is in line with the previous conclusion. 
Similarly, we find that just as in the case of entanglement harvesting, the energy gap also affects the sensitivity of Lorentz violation to mutual information harvesting. 
Since mutual information is the total of classical and quantum correlations, it can be qualitatively assessed that the Lorentz violation inhibits quantum correlations while enhancing classical correlations, and it has a stronger effect on classical correlations.

\begin{figure}[h]
\centering
\includegraphics[scale=0.6]{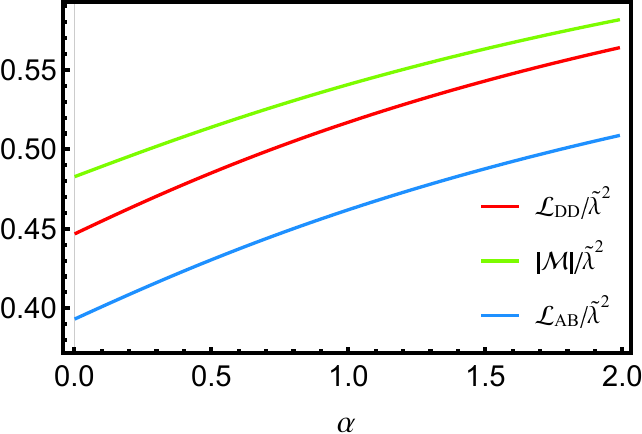}
\caption{ The transition probability $\mathcal{L}_{\mathrm{DD}}$ and the nonlocal correlation terms $\mathcal{|M|}$ and $\mathcal{L}_{\mathrm{AB}}$ are plotted as a function of the Lorentz-violating parameter $\alpha$ with $\Omega\sigma{=}0.01$, $ d_{\mathrm{D}}/\sigma $ = 15, and $ d_{\mathrm{AB}}/\sigma $ = 1.}
\label{PDXC}
\end{figure}

In order to clearly evaluate how the coupling between BTZ black hole spacetime and the Lorentz-violating vector field affects harvested entanglement and harvested mutual information, we illustrate the behavior of the transition probability $\mathcal{L}_{\mathrm{DD}}$, the correlation term $\mathcal{|M|}$ and the $\mathcal{L}_{\mathrm{AB}}$ as a function of the Lorentz-violating parameter $\alpha$ in Fig. \ref{PDXC}. 
\noindent
Indeed, Eqs. (\ref{LDD}), (\ref{eq56}), and (\ref{LAB}) clearly show that as the Lorentz-violating parameter $\alpha$ grows, the transition probability $\mathcal{L}_{\mathrm{DD}}$ and the correlation terms $\mathcal{|M|}$ and $\mathcal{L}_{\mathrm{AB}}$ also increase. 
But the rate of increase is different. 
As shown, the transition probability $\mathcal{L}_{\mathrm{DD}}$ grows significantly faster than the association term $\mathcal{|M|}$. 
And, since concurrence is a competition between these terms, as defined in Eq. (\ref{eq58}), the preceding analysis explains why extracting entanglement reduces with the increases of Lorentz-violating parameter. 
However, because Eq. (\ref{eq59}) is more complex, using a local analysis to solve mutual information is not meaningful.
Nevertheless, based on the previous work, we know that the presence of Lorentz violation helps increase the amount of mutual information harvesting.

\section{CONCLUSIONS AND OUTLOOKS}\label{sec5}

 By performing an appropriate coordinate transformation, the Wightman function in  $\text{AdS}_3$ can be extended to Lorentz-violating regimes. Subsequently, an image-sum approach yields the corresponding Wightman function for the static Lorentz-breaking BTZ-like black hole. Utilizing this Wightman function, we investigate the phenomenon of correlations harvesting between two UDW detectors in the presence of a Lorentz-breaking vector field coupled to the BTZ black hole.

In the context of entanglement harvesting, Lorentz-violating effects exhibit a suppressive influence on the extractable quantum entanglement. However, this suppression contrasts sharply with the enhanced mutual information harvesting in Lorentz-violating BTZ-like spacetime. Given that mutual information quantifies the total correlations, including both classical and quantum components, this highlights the marked disparity in Lorentz-violating effects on quantum and classical correlations: quantum entanglement (e.g., concurrence) is significantly suppressed under Lorentz symmetry breaking, while classical correlations may be amplified due to enhanced field-mode couplings or increased decoherence resistance within modified spacetime structures. This phenomenon suggests that the increase in total correlations within Lorentz-violating backgrounds is predominantly driven by classical components, with quantum correlations experiencing a diminished relative weight in the overall mutual information. This reveals that Lorentz violation, as a quantum property of spacetime, may impose intrinsic constraints on the quantum information capacity encoded in spacetime through resource competition among quantum degrees of freedom.

Notably, due to the inherent properties of black holes, such as Hawking temperature and gravitational redshift factors, entanglement harvesting has a finite capture range defined by the detector's distance from the event horizon. At distances smaller than this range, entanglement ``sudden death" occurs. Unlike entanglement harvesting, mutual information can be extracted at any distance without a capture range limit. Interestingly, Lorentz violation in spacetime extends the range within which entanglement ``sudden death" occurs. This extension also reflects the disruptive influence of spacetime symmetry breaking on quantum entanglement.

The analytical and numerical methods used in this paper can be extended to theoretical models in which black holes couple to Lorentz-violating tensor fields, such as the Kalb-Ramond field \cite{Kalb:1974yc} and various extensions \cite{Yang:2023wtu,Duan:2023gng,Liu:2024oas,Liu:2024lve,Guo:2023nkd,Junior:2024vdk,Jha:2024xtr,Filho:2023ycx,Du:2024uhd,Filho:2024kbq,AraujoFilho:2024rcr,Filho:2024tgy,Hosseinifar:2024wwe}, as well as loop quantum gravity \cite{Brahma:2020eos,Lewandowski:2022zce,Zhang:2024khj,Zhang:2024ney,Liu:2024soc,Liu:2024iec} and modified gravity \cite{Moffat:2014aja,Moffat:2013sja,Liu:2023uft,Liu:2024lbi}.
In such models, extra degrees of freedom in spacetime result in a spacetime geometry significantly different from the metric expression of the BTZ black hole coupled to a vector field, e.g., Eq. \meq{ds2}.
The key mechanism is that extra field coupling alters the event horizon location, causing deviations from the standard BTZ solution. Investigating the correlations harvesting phenomenon arising from vector and tensor couplings in various spacetimes provides deeper insights into the quantum nature of spacetime.

\acknowledgments
We sincerely thank the anonymous referee for their thorough review and helpful suggestions, which have improved the presentation of this work.
The authors gratefully acknowledge Shu-Min Wu for insightful discussions.
This work was supported by the National Natural Science Foundation of China under Grants No.12475051, No.12374408, and No.12035005; the science and technology innovation Program of Hunan Province under grant No. 2024RC1050; the innovative research group of Hunan Province under Grant No. 2024JJ1006; the Natural Science Foundation of Hunan Province under grant No. 2023JJ30384; and the Hunan provincial major Sci-Tech program under grant No.2023ZJ1010.

\appendix
\section{The branch cut problem in $ \mathcal{L}_\text{AB} $}\label{AppendixA}
\begin{figure}[h]
\centering
\includegraphics[scale=0.4]{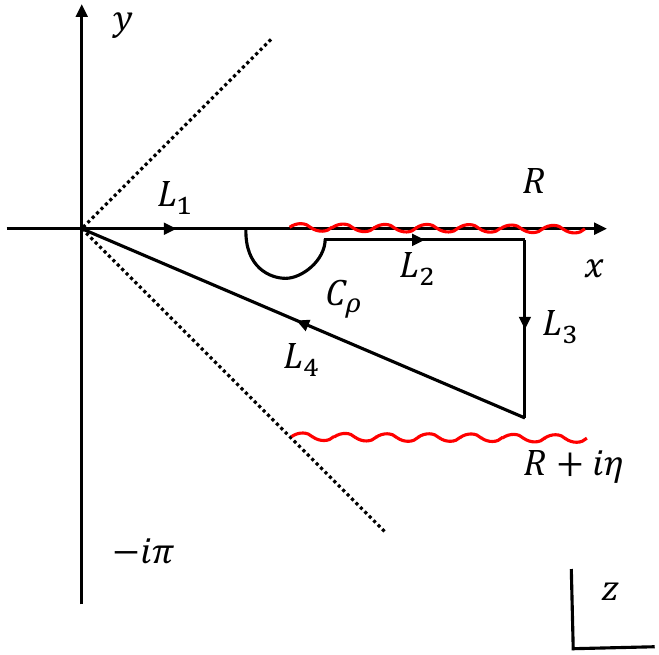}
\caption{ Contour in a complex plane. 
The selected contour needs to lie inside \( y = \pm x \), as indicated by the dashed lines. 
The red wavy lines denote the branch cuts.}
\label{brach1}
\end{figure}

Here, we analyze the branch cut problem associated with $ \mathcal{L}_\text{AB} $, although Eq. (\ref{LAB}) does not have poles, attention still needs to be paid to the branch cuts, which is illustrated in Fig. \ref{brach1}.
Eq. (\ref{LAB}) can be obtained at  $ R \to \infty $ and $\rho \to 0 $ by integrating along paths $ L_1 $, $ C_\rho $, and $ L_2 $.
Here, $\rho$ denotes the radius of the semicircle around the branching point. 
By picking an integration path in the complex plane and utilizing the Cauchy integral theorem, one can perform the complex integral given in Eq. (\ref{LAB}), as
\begin{equation}
\begin{aligned}
\oint_C dz \, e^{-az^2}e^{-i\beta z}
\left[\left(\cosh\chi^{-}_\text{AB,n}-\cosh z\right)^{-1/2}\right.\\
\left.-\zeta\left(\cosh\chi^{+}_\text{AB,n}-\cosh z\right)^{-1/2} \right].
\end{aligned}
\end{equation}
When we set  $ z = x + iy $ and assume that $\Omega > 0$, the conditions under which the chosen contour leads to convergence, as required by Eq. (\ref{LAB}) is  $ \left| e^{-a z^2} e^{-i \beta z} \right| = e^{-a (x^2 - y^2)} e^{\beta y} < 1 $, thus, the contour $ C $ should satisfy $ -x \leq y \leq x $  and $ y < 0 $. Consequently, by choosing the contour $ C = L_1 C_\rho L_2 L_3 L_4 $ in the figure and using the Cauchy integration theorem as well as the fact that $ \lim_{R \to \infty} \int_{L_3} = 0 $ holds, Eq. (\ref{LAB}) evolves as follows
\begin{equation}
\begin{aligned}
\mathcal{L}_{\text{AB}} =& 2K \sum_{n=-\infty}^{\infty} \lim_{R \to \infty} \operatorname{Re} \int_{0}^{R+i \eta} dz e^{-az^2} e^{-i\beta z} \\
&\times\left[\tfrac{1}{\sqrt{\cosh\chi^{-}_\text{AB,n}-\cosh z}}-\tfrac{\zeta}{\sqrt{\cosh\chi^{+}_\text{AB,n}-\cosh z}} \right],
\end{aligned}
\end{equation}
where $ \eta \in (-\pi, 0)  $.

%

\end{document}